\documentclass{aa}
\usepackage{graphicx}
\usepackage{txfonts}
\usepackage{epstopdf}
\usepackage{natbib}
\bibpunct{(}{)}{,}{a}{}{;}
\begin{document}
\title{Disk of the Small Magellanic Cloud as traced by Cepheids}
\author{Smitha Subramanian\inst{1},
Annapurni Subramaniam\inst{1}}
\institute{Indian Institute of Astrophysics, Koramangala II Block, Bangalore-560034, India\\
	              \email{smitha@iiap.res.in, purni@iiap.res.in}}
\date{Received, accepted}
\abstract
{The structure and evolution of the disk of the Small Magellanic Cloud (SMC) are traced by studying the Cepheids.}  
{We aim to estimate the orientation measurements of the disk, such as the inclination, $i$, 
and the position angle of the  line of nodes (PA$_{lon}$), $\phi$, and the depth  
of the disk. We also derive the age of the Cepheids and hence the age distribution of the SMC Cepheids.}
{We used the {{\it V}} and {{\it I}} band photometric data of the fundamental and first-overtone Cepheids 
from the Optical Gravitational Lensing Experiment survey. The period-luminosity (PL) 
relations were used to  
estimate the relative distance and reddening of each Cepheid. 
The Right Ascension, Declination, 
and relative distance from the centroid of each Cepheid were converted 
into x, y, and z Cartesian coordinates. A weighted least-square plane fitting method was then applied to 
estimate the structural parameters. The line-of-sight depth  
 and then the orientation corrected depth or thickness of the disk were estimated from the relative 
distance measurements.  
The period-age-colour (PAC) relation of Cepheids were used to derive the age of the 
Cepheids.}
{A break in the PL relations of both the fundamental mode and first-overtone Cepheids 
at P $\sim$ 2.95 days and P $\sim$ 1 day are observed. An inclination of 64$^o$.4$\pm$0$^o$.7 
and a PA$_{lon}$ = 155$^o$.3$\pm$6$^o$.3 are obtained from the full sample. 
A reddening map of the SMC disk is also 
presented. The orientation-corrected depth or thickness of the SMC disk is found to be 1.76 $\pm$ 0.6 kpc. 
The scale height is estimated to be 0.82 $\pm$ 0.3 kpc. The age distribution of Cepheids matches the SMC 
cluster age distribution.}
{The radial variation of the disk parameters mildly indicate 
structures/disturbances in the inner SMC (0.5 $<$ r $<$ 2.5 degree). 
Some of the Cepheids found in front of the fitted plane in the eastern regions are possibly the youngest 
tidally stripped counterpart of the  H {{\sc i}} gas of the Magellanic Bridge. 
The Cepheids behind the 
fitted plane are most likely the population in the Counter Bridge predicted in recent numerical 
simulations. Different scenarios for the origin of the extra-planar Cepheids are also discussed.}
 
\authorrunning{Subramanian \& Subramaniam}
\titlerunning{Disk of the Small Magellanic Cloud}
\maketitle
\section{Introduction}
The Small Magellanic Cloud (SMC), located at a distance of around 60 kpc, is one of the nearest galaxies. 
The SMC is believed to have had interactions with the Large Magellanic Cloud (LMC) and also with the Milky Way. 
Features such as the Magellanic Bridge and the Magellanic Stream are considered as the observational 
signatures of these interactions.
Recent proper motion estimates and related studies (\citealt{k13}, \citealt{k06a}, \citealt{k06b} and 
\citealt{bk07}) indicate that the LMC and the SMC (together known as the Magellanic Clouds, MCs) are 
approaching our Galaxy for the first time. These results also claim that the MCs might not have always 
been a binary system. \cite{bes10} demonstrated that the Magellanic Bridge and the Magellanic 
Stream are formed by the mutual interaction of the MCs before they have been accreted by the Milky Way. 
\cite{bc08} suggested that the SMC 
might have undergone a major merger event in the early stage of its evolution. Therefore, the structure of the 
SMC might have been modified by the merger events in its early evolution and in the recent past by  
the interactions with the LMC. These interactions might have also affected the SFH  of the SMC. 

Both theoretical and observational studies (\citealt{gn96}, \citealt{bc08}, \citealt{z00}, 
\citealt{hz06}, 
\citealt{eh08}) indicate that the SMC is a two-component system where old and intermediate-age 
stars are distributed in a spheroidal or slightly ellipsoidal component and young stars and gas 
are distributed in a disk. From studying the old RR Lyrae stars and the intermediate-age red clump stars in the SMC, \cite{ss12} 
found that both these populations have 
a slightly ellipsoidal distribution and are located in a similar volume. They also estimated structural 
parameters of the ellipsoidal component of the SMC. \cite{has12a} studied the RR Lyrae stars in the SMC and 
obtained similar results. 

Young stars (age $<$200 Myr) and the H {\sc i} gas are good 
tracers to understand the disk properties. The analysis of young stars (age $<$ 200 Myr) \citep{z00} and 
high-resolution H {\sc i} observations \citep{stan04} show that the SMC disk is   
quite irregular and asymmetric. The H {\sc i} observations also show that the SMC has a 
significant amount of rotation with a circular velocity of approximately 60 kms$^{-1}$ \citep{stan04} and a 
steep velocity gradient of 91 kms$^{-1}$ in the south-west to 200 kms$^{-1}$ in 
the north-east. \cite{eh08} obtained velocities for 2045 young (O, B, A) stars
in the SMC and found a velocity gradient of similar slope as seen in the H {\sc i} gas. 
Surprisingly, however they found a position angle  ($\sim$ 126$^o$)  for the line of 
the steepest velocity gradient that is quite different, and almost orthogonal to that seen 
in the H {\sc i}. Using Cepheid data, 
\cite{cc86} found the SMC to consist of a central bar seen edge-on, 
a near arm in the north-east (NE), a far arm in the south-west (SW), and a mass of 
material pulled out of the centre and seen in fornt of the SW arm. 
They obtained an inclination, $\it{i}$ = 70$^\circ$.0$\pm$3$^\circ$.0, and a position 
angle of the closest part of 58$^\circ$$\pm$10$^\circ$ from studying  
63 Cepheids. \cite{ls86} obtained $\it{i}$ = 45$^\circ$ $\pm$ 7$^\circ$ and 
a position angle of the closest part to be 55$^\circ$ $\pm$ 17$^\circ$. From the study of 
Cepheids in the bar region of the SMC, 
\cite{g00} estimated an inclination of 68$^\circ$.0$\pm$2$^\circ$.0 
and position angle of the line of nodes, $\phi$, of 238$^\circ$.0$\pm$7$^\circ$.0. 

Another interesting feature of the SMC is its high line-of-sight (LOS) depth. \cite{mf86} \& \cite{mf88}  
found from studying SMC Cepheids that the SMC has a considerable line-of-sight depth of $\sim$ 20 Kpc. 
\cite{c01} estimated a depth of 6-12 Kpc from studying clusters. From the analysis of red 
clump stars and RR Lyrae stars, \cite{ss12} 
found an LOS depth of 14 kpc, 
which corresponds to the $\sim$ 3$\sigma$ value. \cite{Kapakos11} estimated an 
LOS depth of 4.13$\pm$ 0.27 kpc, which corresponds to the 1$\sigma$ depth from the study of RR Lyrae stars.  
For old and intermediate-age stars in the SMC, a high LOS depth is expected because they 
are assumed to be distributed in a spheroidal or slightly ellipsoidal 
system. For the young stars such as the Cepheids, a high LOS depth is likewise expected as they are distributed in a highly 
inclined plane. The LOS depth has to be corrected for the orientation of the disk to obtain the depth of the disk. 

\cite{has12a} (hereafter H12) studied the fundamental-mode Cepheids in the SMC identified by the 
Optical Gravitational Lensing Experiment (OGLE III) survey. 
They estimated an inclination of 74$^\circ$ $\pm$ 9$^\circ$. From the number density map, they also 
estimated the position angle of the major axis of the SMC disk, which is given as 66$^\circ$ $\pm$ 15$^\circ$. 
The individual distances of Cepheids obtained 
in H12 range from 45-75 kpc, which corresponds to a wide extent in the line-of-sight depth. 
They estimated the LOS depth by dividing the observed regions into sub-regions and taking the dispersion 
in the distances (corrected for uncertainties in distances) in the sub-regions. From the upper and lower 
panels of Fig.6 of H12 the LOS depth values range from $\sim$ 3-9 kpc. They also 
estimated a LOS depth of 4.2 $\pm$ 0.4 kpc (corresponding to 1 $\sigma$ depth) from studying RR Lyrae stars. 
It was found that the scale height of Cepheids exceeds that of the 
RR Lyrae stars. Scale height of younger population is commonly expected to be lower than that of older 
population. The greater scale height for the Cepheids obtained by H12 compared with the older 
population may be due to the fact the LOS depth is not corrected for the inclination. 

In the present study the $\it{VI}$ band photometric data of the fundamental and first-overtone mode Cepheids 
in the SMC, identified from the OGLE III survey, are 
analysed to understand the structural parameters, orientation-corrected depth/thickness, and the star 
formation history of the SMC disk. H12 derived the inclination, position angle of the major axis, and the LOS depth 
of the SMC disk using only the fundamental-mode Cepheids, whereas we study both the fundamental and first-overtone Cepheids, which 
makes the sample size larger. H12 used a single period - luminosity (PL) relation given by \cite{sandage09} to derive 
the distances to each Cepheid. On the other hand, because we identified a break in the PL relations of both the fundamental and first over 
tone sample, the separate PL relations of the shorter and longer period sample were used to derive the relative 
distances and hence the structural parameters of the SMC disk. This improves the accuracy of the estimated 
parameters. Along with the inclination, position angle of the major axis, and the LOS depth, we also derived 
the position angle of the line of nodes and the orientation-corrected scale height of the SMC disk. We also  
estimate the age of the individual Cepheid and hence the age distribution of the Cepheids in the  
SMC disk. A detailed comparison of the study of H12 with the present work is given in Sect. 5.2.1.
The structural parameters we estimate are the inclination, $\it{i}$ , the position angle 
of the line of nodes, $\phi$, and the line-of-sight depth. The PL relations are obtained 
and are used to estimate the structural parameters of the SMC disk. As a by-product of 
this study, a reddening map towards the SMC disk is also obtained. 
Cepheids also obey period-age (PA) and period-age-colour (PAC) relations which can be used to derive the star formation history 
(SFH) of the external galaxies. Here we use the PAC relation to estimate the age and hence try to 
understand the SFH of the SMC disk. 

The structure of the paper is as follows: The next section describes the optical data of the 
Cepheids. Sect. 3 reports the methodology involved in estimating the structural parameters and the 
SFH of the SMC disk. The results are presented in Sect. 4. The discussion and summary are given in Sects. 5 and 6.  
\section{Data}
The seventh part of OGLE III catalog \citep{u10smcc} of variable stars consists of 4630 
classical Cepheids in the SMC. In this catalogue 2626 classical Cepheids are fundamental-mode pulsators, and 
1644 Cepheids are first overtone pulsators. 
Among the total 4270 Cepheids taken for the study, only 4235 (2603 fundamental mode and 1632 
first-overtone) 
stars have simultaneous detections 
in the V and I bands, and we considered only these stars for our analysis. 

\section{Methodology}

\subsection{Structural parameters}

The PL relation is used to estimate the relative distance of 
each Cepheid from the centre of the SMC. The methodology is described below. 
It is similar to that used by \cite{n04} for the analysis of LMC Cepheids. 
The PL relation is given by
\begin{equation}
\overline{M_\lambda} = \alpha_{\lambda}\log(P) + \beta_{\lambda} + \epsilon_{\lambda} (M, T_{eff}, Z),
\label{1}
\end{equation}
where $\lambda$ denotes the photometric bands in which Cepheids are observed, $\overline{M_\lambda}$ 
is the mean intrinsic magnitude, $\alpha$ and $\beta$ are the PL coefficients and 
P the pulsation period. $\epsilon$ is some unknown function of stellar parameters, such as
pulsation mass, effective temperature, and metallicity.  
Equation (1) is exact, because function $\epsilon$ includes all parameters that affect the stellar luminosity.
The $\overline{M_\lambda}$ can be converted into observed mean magnitude, 
$\overline{m_\lambda}$ using the equation
\begin{equation}
 \overline{m_\lambda} = \mu + A_{\lambda} + \alpha_{\lambda}\log(P) + \beta_{\lambda} + \epsilon_{\lambda} ,
\label{2} 
\end{equation}
where $\mu$ and A$_{\lambda}$ are the distance modulus and extinction in the photometric band denoted 
by $\lambda$. A$_{\lambda}$ = R$_{\lambda}$E($B-V$), where R$_{\lambda}$ 
is the ratio of total to selective extinction and E($B-V$) is the reddening. R$_{V}$ is 3.12 
\citep{bb88}. R$_{I}$ is taken as 1.75, which is calculated 
as follows: E$(V-I)$ = 1.25 E$(B-V)$ \citep{bb88} and A$_I$ = 1.4 E$(B-V)$ \citep{sch98}.
From these two equations we can find that A$_I$ = 1.75 E$(B-V)$. 

The general form of function $\epsilon$ is unknown, and as given in \cite{n04},    
we assume $\epsilon$ has a distribution with mean zero and dispersion $\sigma_{0}$. The main effect of
of the unknown physics is on the dispersion of the observed PL relation. Note that the dispersion is in 
general a function of wavelength. There is no loss of generality in assuming zero
mean for the distribution of $\epsilon$, since zero point $\beta$ can always
be shifted to accommodate any mean $\epsilon$.  

The quantities $\mu$ and 
E($B-V$) can be divided into two parts, one mean quantity corresponding to the entire galaxy, and the other 
varies from star to star. The index $\it{i}$ in the equations below denotes an individual star.
\begin{equation}
 \mu_i = \overline{\mu}_{galaxy} + \delta\mu_i
\label{3}
\end{equation}
\begin{equation}
 E(B-V)_i = \overline{E(B-V)}_{galaxy} + \delta E(B-V)_i 
\label{4}
\end{equation}
The mean distance modulus, the mean extinction, and the mean of the function $\epsilon$ (here we have assumed  
$\epsilon$ as a function with mean zero) can be incorporated 
into the quantity $\beta$ and the equation for each star 
connecting the mean observed magnitude ($\overline{m_{\lambda,}}_{i}$), relative distance,  
and relative reddening is given as
\begin{equation}
\overline{m_{\lambda,}}_{i} = \delta\mu_i + R_{\lambda}\delta E(B-V)_i + \alpha_{\lambda}\log(P)_i + 
\beta^{'}_{\lambda} ,
\label{5}
\end{equation}
 

The PL relation coefficients ($\alpha_{\lambda}$ and $\beta^{'}_{\lambda}$) of the sample using 
the observed mean V and I magnitudes are obtained. These coefficients are substituted into the PL relation 
given by Eq. 5. 
The variation in distance modulus and reddening of each star from the mean values 
towards the SMC are estimated by solving the equations for V and I passbands. The errors assosicated 
with the relative distance modulus and the relative reddening are estimated by considering the fit errors in the 
co-efficients of the PL relations, the photometric errors, and the intrisic dispersion of the PL relation. 
\cite{Gieren98} estimated the intrisic dispersion in the LMC Cepheid PL relation to be $\sim$ 0.11 mag 
and also found that the value remains more or less the same from V to K band. Under the assumption that the 
intrisic dispersion in the Cepheid PL relation of the SMC is similar to that of the LMC, we used the value of 
0.11 mag as the intrisic dispersion (in V and I bands) in estimating the errors associated with the 
realtive distances of our sample. Again, the final values of the structural parameters are not 
very sensitive to the choice of this value \citep{n04}. 

The relative distance modulus 
$\delta$$\mu_i$ is converted into relative distance, $\Delta$D by the equation 
$\Delta$D$_i$ = D$_0$(10$^{\delta \mu_i /5}$ -1). D$_0$ is the mean distance to the SMC. 
The error in relative distance is also calculated using the error propagation rule for exponentiation. The 
application of the 
error propagation rule indicates that the errors are large for Cepheids at farther distances. 
The RA, Dec, and the relative distance are converted into x,y,z coordinates using the transformation 
equations given below (\citealt{vc01}, see also the Appendix of \citealt{wn01}).
\begin{equation}
x = -D\sin(\alpha - \alpha_0)\cos\delta
\label{6}
\end{equation}
\begin{equation}
 y = D\sin\delta\cos\delta_0 - D\sin\delta_0\cos(\alpha - \alpha_0)\cos\delta
\label{7}
\end{equation}
\begin{equation}
z = D_0 - D\sin\delta\sin\delta_0 - D\cos\delta_0\cos(\alpha - \alpha_0)\cos\delta ,
\label{8}
\end{equation}

where D, the distance to each Cepheid is given by D = D$_0$ $\pm$ $\Delta$D.
($\alpha$, $\delta$) and ($\alpha_0$, $\delta_0$) represent the RA and Dec of 
each Cepheid and the centroid of the sample. The x-axis is antiparallel to the RA axis, the 
y-axis is parallel to the declination axis, and the z-axis is toward the observer. Because the structural 
parameters of the galaxy are estimated based on the relative 
variation of the distance of the sources within the galaxy, the mean value of distance we take 
does not affect the parameter determination.

After determining the x,y, and  z coordinates, we applied a weighted least-square
plane fit to obtain the structural parameters of the SMC disk. 
The equation of the plane assumed is 
\begin{equation}
     z = Ax + By + C.
\label{9}
\end{equation}

From the coefficients of the planar equation A, B, and C, the structural parameters $\it{i}$ and $\phi$ 
can be calculated using the formula
\begin{equation}
inclination, {i} = \arccos(C/\sqrt{A^2+B^2+1}) 
\label{10}
\end{equation}
\begin{equation}  
PA_{lon}, \phi = \arctan(-A/B)+sign(B)\pi/2.
\label{11}
\end{equation}
The deviations of each Cepheid from the fitted plane can be calculated.  
The expected z for a Cepheid in the plane is estimated from the 
equation of a plane. The difference in the expected and 
calculated z values is taken as the deviation of the Cepheid from the fitted 
plane. Thus the regions with extra-planar features can be identified 
and quantified. When the deviations are estimated, the Cepheids with  
deviations stronger than twice the error in z are omitted and the plane-fitting 
procedure is applied to the remaining regions to re-estimate the 
structural parameters of the SMC disk plane. The error in the estimate
of the SMC disk parameters is calculated by propagating the
systematic errors associated with all the quantities involved in
the estimation. 

The range of relative distances of Cepheids with respect to the mean distance of 
the SMC gives an estimate of the LOS depth of the disk. The depth/thickness of the SMC 
disk can be estimated by correcting the LOS depth for the orientations of the disk with respect to 
the sky plane. The actual thickness of the SMC disk can be estimated after deprojecting  
the observed data to the SMC plane. Once we have the orientation measurements ($\phi$, $\it{i}$), the (x,y,z) 
coordinates in the sky plane can be transformed into the actual 
plane of the SMC (x',y',z') by using the transformation equations given below \citep{vc01},
\begin{equation}
 x' = x\cos(\phi + 90) + y\sin(\phi + 90)
\label{12}
\end{equation}
\begin{equation}
 y' = -x\sin(\phi + 90)\cos\it{i} + y\cos(\phi + 90)\cos\it{i} -z\sin\it{i}
\label{13}
\end{equation}
\begin{equation}
 z' = -x\sin(\phi + 90)\sin\it{i} + y\cos(\phi + 90)\sin\it{i} +z\cos\it{i}.
\label{14}
\end{equation}
The range of z' gives the measure of the actual depth/thickness of the SMC disk. 

\subsection{Age of the Cepheids}
The period - age (PA) relation of Cepheids are used in general to estimate the age of each Cepheid and 
hence the recent SFH of the SMC.  
\cite{efr78} derived an empirical PA relation by adopting Cepheids in Galactic, in M31, and in 
LMC clusters, whose age was estimated independently. \cite{mg97} derived a new semi-empirical PA relation 
and used Cepheids in NGC 206,
the super association in M31, to trace the age distribution and, in turn, the SFH in 
this region, located at the intersection of two spiral arms. A similar approach was also
adopted by \cite{efel98}, \cite{grbr98}, and \cite{efr03}, who also derived 
new empirical PA relations on the basis of a larger sample of cluster Cepheids. \\
Later, \cite{bono05} presented 
new theoretical PA and PAC relations for the fundamental and first overtone Cepheids for 
accurate individual stellar age estimations in our Galaxy and in the MCs. These theoretical relations 
were based on homogeneous and detailed sets of nonlinear, convective pulsation models covering a 
broad range of stellar masses and chemical compositions, together with 
evolutionary models. They indicated that 
using the PAC relation improves the accuracy of age estimates in the long-period 
(log P $>$ 1) range (young Cepheids), since they account for the position of individual objects inside the 
instability strip. The PAC relation for the fundamental mode and first-overtone 
Cepheids presented by \cite{bono05} for the SMC metallicity of (Z=0.004) are 
\begin{equation}
 \log(t) = 8.24\pm0.09 - 0.88\pm0.02log(P) + 0.42\pm0.08(V-I)_0
\label{15}
\end{equation}
and
\begin{equation}
 \log(t) = 8.06\pm0.06 - 1.16\pm0.02log(P) + 0.64\pm0.08(V-I)_0,
\label{16}
\end{equation}
where t is the age of the Cepheid in 
years, P is the period in days, and $(V-I)_0$ is the intrinsic colour of the Cepheid. \cite{bono05} 
performed a detailed comparison between 
the evolutionary ages, based 
on isochrone fit and pulsation ages based on the PA and PAC relations of a sample of LMC and SMC clusters 
with at least two Cepheids. The difference in evolutionary and pulsation ages was found to be smaller than 
20\%. Based on these results, \cite{bono05} proposed that the PA and PAC realtions can be used for 
accurate age estimations and hence to trace the age distribution across the MCs. Later, \cite{mar2006} 
applied these relations to Cepheids in the inner regions of the MCs to constrain the SFH  
of these galaxies. In our analysis, we used the PAC relations given by \cite{bono05} for the SMC metallicity to 
estimate the age distribution of our sample Cepheids.

\section{Results}
\subsection{PL relations}
A break in the PL relation of the SMC 
fundamental-mode Cepheids at about 2.5 days was reported by \cite{bauer99} and confirmed by \cite{u99cep}, 
\cite{sharpee02}, \cite{sandage09}, and \cite{u10smcc}. From the detailed 
analysis of Cepheids in the bar region of the SMC, \cite{tam11} found that the PL relation 
of fundamental-mode Cepheids has a break at logP = 0.55 (which approximately corresponds to a period 
of 3.548 days) and the PL relation of first-overtone Cepheids has a break at logP = 0.4 
(which corresponds to a Period of 2.5 days). Other than in the study of \cite{tam11},  
linear regression analysis was not performed to estimate the break points. 

We analysed our sample independently using two linear regression model, where the 
break point is considered as a free parameter. A small discontinuity at the break point 
is also considered as a requirement for the fit. Thus the break points, which minimize the 
residuals of the fits in the fundamental-mode and 
first-overtone Cepheids, are estimated separately. The PL 
diagrams with the break points are shown in Fig. 1. The PL relations of fundamental-mode Cepheids 
in V and I bands have a break at log(P) = 0.47, which corresponds to a period of $\sim$ 2.95 days. The 
PL relations of first-overtone Cepheids show a break at log(P) = 0.029, which corresponds to a period of 
$\sim$ 1 day. \cite{kan06} investigated the break in the LMC PL relation (at P = 10 days) and proposed that the break  
might be the result of interaction of the hydrogen ionization front with the photosphere of the star. They also mentioned that 
the metallicity of the host galaxy and the sample affect the interaction of the hydrogen ionization front with the photosphere and 
hence in turn affect the break in PL relations. Based on 
stellar evolution, \cite{cor03} claimed that the fainter Cepheids in the SMC are metal poor. The metal-poor nature of the fainter 
Cepheids may also be the reason for the observed break in PL relation, and this possibility is discussed in Sect. 5.1. 
Based on the observed break in the PL relations, we divided the fundamental-mode and first-overtone 
data set into two sub-groups each.

The 2603 fundamental mode Cepheids were divided into one group with log(P)$>$ 0.47 and 
 an other group with log(P)$<$ 0.47. The first group contains 789 stars, latter group 
contains 1814 stars. 
The PL relations 
for the two groups were obtained separately using least-square fit of the observed magnitudes and 
periods with 3$\sigma$ clipping. After applying 3$\sigma$ clipping to the data in the V and I bands, we had 778 stars in the 
longer period group and 1764 stars in the shorter period group 
for further analysis in the fundamental-mode Cepheids. 
The slope, intercept, and standard deviation of the best fit are given in Table. 1. 
The lower left and upper left panels of Fig. 1 show the V and I 
band PL diagrams of fundamental-mode Cepheids with the break points and best fits. 
From the plots  and table we can see that the slope of the sample with shorter 
period Cepheids is steeper than that of the longer period Cepheids. 

\begin{figure}
\includegraphics[width=9.3cm,height=10cm]{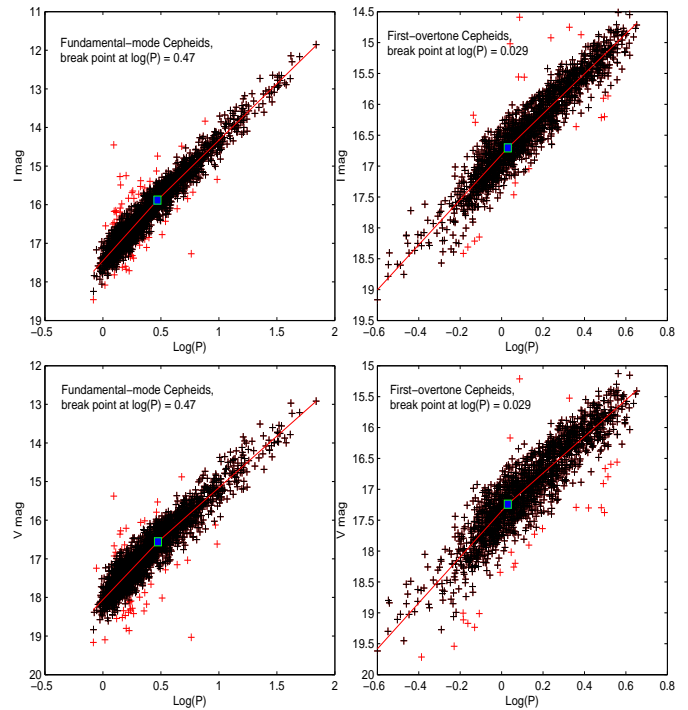}
\caption{V and I band PL diagrams of fundamental-mode and first-overtone Cepheids. The estimated break points 
are also shown as blue squares. 
The red points 
in all the panels represent the outliers after a 3$\sigma$ clipping was applied.}
\end{figure}

\begin{table*}
\centering
 \caption{Coefficients of PL relations of fundamental-mode and first-overtone Cepheids}
\label
{Table:1}
\vspace{0.25cm}
\begin{tabular}{lrrr}
\hline
 Data & $\alpha_{\lambda}$ & $\beta^{'}_{\lambda}$ & $\sigma$\\
\hline
\hline
{\bf Fundamental-mode Cepheids}\\
\hline
{\bf This study, break point at log(P) = 0.47}\\
V-band data with log(P) $>$ 0.47 &-2.67$\pm$0.041 & 17.82$\pm$0.032 & 0.28\\
V-band data with log(P) $<$ 0.47 &-3.19$\pm$0.053 & 18.06$\pm$0.013& 0.26\\
I-band data with log(P) $>$ 0.47 &-2.93$\pm$0.032 & 17.26$\pm$0.025 & 0.22 \\
I-band data with log(P) $<$ 0.47 &-3.37$\pm$0.043 & 17.47$\pm$0.011 & 0.21\\
\hline
{\bf \cite{tam11}, break point at log(P) = 0.55. Adopted (m-M)$_0$ = 18.93.}\\
V-band data with log(P) $>$ 0.55 &-2.53$\pm$0.056 & -1.466$\pm$0.050 & 0.25\\
V-band data with log(P) $<$ 0.55 &-3.20$\pm$0.060 & -1.071$\pm$0.018& 0.22\\
I-band data with log(P) $>$ 0.55 &-2.84$\pm$0.043 & -1.872$\pm$0.038 & 0.19 \\
I-band data with log(P) $<$ 0.55 &-3.37$\pm$0.046 & -1.577$\pm$0.013 & 0.17\\
\hline
{\bf First-overtone Cepheids}\\
\hline
{\bf This study, break point at log(P) = 0.029}\\
V-band data with log(P) $>$ 0.029 &-3.03$\pm$0.056 & 17.33$\pm$0.016 & 0.27\\
V-band data with log(P) $<$ 0.029 &-3.59$\pm$0.125 & 17.35$\pm$0.017& 0.30\\
I-band data with log(P) $>$ 0.029 &-3.23$\pm$0.045 & 16.79$\pm$0.013 & 0.22 \\
I-band data with log(P) $<$ 0.029 &-3.63$\pm$0.101 & 16.81$\pm$0.014 & 0.25\\
\hline
{\bf \cite{tam11}, break point at log(P) = 0.4. Adopted (m-M)$_0$ = 18.93.}\\
V-band data with log(P) $>$ 0.4 &-2.53$\pm$0.056 & -2.028$\pm$0.026 & 0.24\\
V-band data with log(P) $<$ 0.4 &-3.20$\pm$0.060 & -1.759$\pm$0.010& 0.25\\
I-band data with log(P) $>$ 0.4 &-2.84$\pm$0.043 & -2.423$\pm$0.020 & 0.19 \\
I-band data with log(P) $<$ 0.4 &-3.74$\pm$0.046 & -2.210$\pm$0.008 & 0.19\\
\hline
\end{tabular}
\end{table*}

The sample of 1632 first-overtone Cepheids were also divided into two sub-groups. 
There are 569 first-overtone Cepheids which have log(P) $<$ 0.029 and 1063 Cepheids 
with log(P) $>$ 0.029. The PL relations 
for the two groups obtained separately using least-square fit of the observed magnitudes and 
periods with 3$\sigma$ clipping. After applying 3$\sigma$ clipping to the data in V and I bands, we had 
1041 stars in the longer period group and 559 stars in the shorter period group 
for further analysis in the first-overtone mode Cepheids. The slope, intercept, and standard deviation 
($\sigma$) of the best 
fits are given in Table. 1. The lower right and upper right panels of Fig. 1 show the V and I 
bands PL diagrams of first-overtone Cepheids with the break points and best fits. 
As for the fundamental-mode Cepheids, the first-overtone sample shorter Cepheids also 
have steeper slope than the longer period Cepheids. 

For the fundamental-mode Cepheids, the break point estimated by \cite{tam11} 
is close to what we have obtained. They had also divided the fundamental-mode sample into 
two sub-groups based on the break point they estimated. The PL co-efficients obtained for 
the two sub-groups by \cite{tam11} are given in Table. 1. 
From the table, we can see that the slope values 
they estimated for the two sub-groups of their fundamental mode Cepheids match with our estimates well.
On the other hand, the break point and the PL coefficients estimated for the first-overtone 
Cepheids in their study differ from that of our estimates. The difference in the sample size of the two studies 
might cause a difference in the estimates of the prominent breaks. The probable reasons for the differences in 
the two studies are discussed in Sect. 5.1.

\begin{table*}
\centering
\caption{Centroid coordinates and position angle of the major axis of different sub-groups in the Cepheid sample.}
\label{Table:2}
 \begin{tabular}{lllll}
\hline
\hline
Sample & Number of & RA of the centroid  & Dec of the centroid&Position angle of\\
 & Cepheids& (hh:mm:ss) & (degree)&Major axis\\
\hline
Fundamental-mode Cepheids with log(P) $<$ 0.47  &1764&00:54:41 & -73.02 & 72$^\circ$.1$\pm$1$^\circ$.0\\
Fundamental-mode Cepheids with log(P) $>$ 0.47  &778&00:56:41 & -72.90& 65$^\circ$.7$\pm$1$^\circ$.5 \\
Fundamental-mode Cepheids &2542&00:55:18 & -72.99 &70$^\circ$.1$\pm$0$^\circ$.8 \\
First-overtone Cepheids with log(P) $<$ 0.029 &559&00:55:11 & -73.06 & 73$^\circ$.3$\pm$1$^\circ$.8\\
First overtone Cepheids with log(P) $>$ 0.029 &1041& 00:56:37& -72.93 & 70$^\circ$.0$\pm$1$^\circ$.3\\
First-overtone Cepheids &1600&00:56:07 & -72.97&71$^\circ$.2$\pm$1$^\circ$.1\\
Fundamental \& first-overtone Cepheids &4142&00:55:37& -72.98&70$^\circ$.5$\pm$0$^\circ$.6\\
\hline
\hline
  ×
 \end{tabular}

\end{table*}

\begin{figure*}
\resizebox{\hsize}{!}{\includegraphics{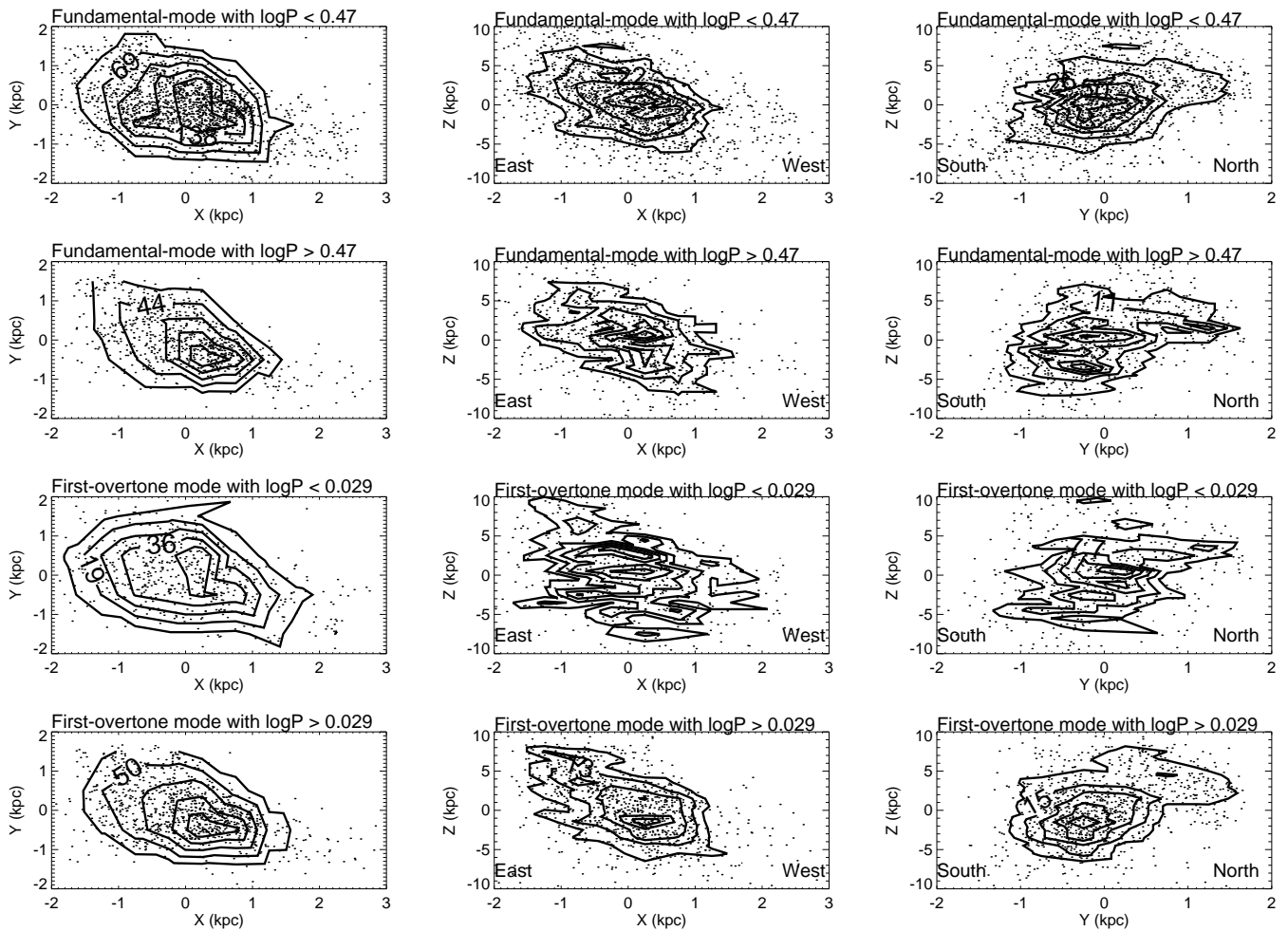}}
\caption{XY, XZ and YZ distributions of fundamental-mode and first-overtone Cepheids. The left 
panels show the XY distributions of different sub-samples.  
The middle panels show the XZ distributions of different sub-samples. The right panels show the YZ 
distributions of different sub-samples. The convention of z-axis is such that the +ve z-axis represents the 
direction towards the observer and the -ve z-axis represents the direction away from us. 
The density contours are overplotted in all the panels. }
\end{figure*}

\subsection{Spatial distribution}
The estimated coefficients were substituted in the respective 
PL relations of V and I bands (Eq. 5) and these  
equations were solved simultaneously to obtain the variation of distance and reddening of each star 
from the mean value. We used the PL relations for shorter and longer period Cepheids in the fundamental-mode and first-overtone 
modes. The RA, Dec, and variation in the distance are converted into x,y,z 
coordinates. 
The z coordinate represents the relative distance of 
the star with respect to the centre of the SMC. The distance towards the SMC centre was taken as 
60 kpc. The convention used is such that the +ve z axis points towards us, the -ve 
z axis away from us. 
We estimated the x,y,z coordinates with respect to the centroid of the different sub-groups 
of the sample.  The centroid values of the different sub-groups of the sample and the 
whole sample are given in Table 2. From the table we can see that the shorter period sample in both the fundamental 
and first-overtone sample have a similar centroid. Similarly, the longer period Cepheids of the fundamental and first-overtone 
sample have similar centroid. The centroid of the longer period Cepheids is shifted towards the 
north-eastern part from that of the shorter period Cepheids. 

The fit error of the PL co-efficients, the uncertainty in the reddening law, and the photometric 
errors were propagated properly to estimate the error in relative magnitude variation and hence the error 
in relative distance. Because the relative distance estimates were used to calculate the values of z, the 
error in the estimation of relative distances was propagated using the error propagation rule 
to quantify the error in z. The average error in z is $\sim$ 2.8 kpc.  

The XY, XZ, and YZ distributions of shorter and longer period sub-samples of the fundamental mode and first 
over tone Cepheids in the SMC are shown in Fig. 2. 
From the XY distributions shown in the left panels, we can see that 
the longer period Cepheids, especially the longer period ones in the fundamental-mode sample, are 
concentrated more along the bar region. We estimated the position angle of the 
major axis of the XY distributions of all the sub-samples. The estimates are given in Table. 2. The values 
in the table are comparable with the estimates by H12.  

From the middle panels (XZ distribution) we can see that the Cepheids 
in the eastern part of the SMC are closer. We can also see a gradient in the relative 
distances of the Cepheids from east to west. The linear drift 
in the relative distances is the effect of the inclination of the SMC disk. 
The plots show that the longer period fundamental-mode Cepheids have 
steeper gradient than the shorter period ones. For the shorter period first-overtone Cepheids, 
the gradient is not very clearly seen. This may be due to the small sample size. 
The longer period first-overtone Cepheids have a steeper gradient than the 
fundamental-mode Cepheids. 

The YZ distributions shown in the 
right panels of Fig. 2 suggest that there is a nearly symmetric distribution of 
Cepheids along the line of sight with respect to the centre in the southern side (Y $<$ 0) of the SMC. 
On the other hand, more stars are closer to us in the northern side. This 
mildly suggest that on an average the northern side of the SMC is closer to us than 
the southern side. This trend is more prominent for the outer contours. 
We do not see any gradient from north to south as seen from east to west. The YZ distribution of shorter period 
first-overtone Cepheids is not very well described; this may again be because of the small sample size. 

\begin{figure*}
\resizebox{\hsize}{!}{\includegraphics{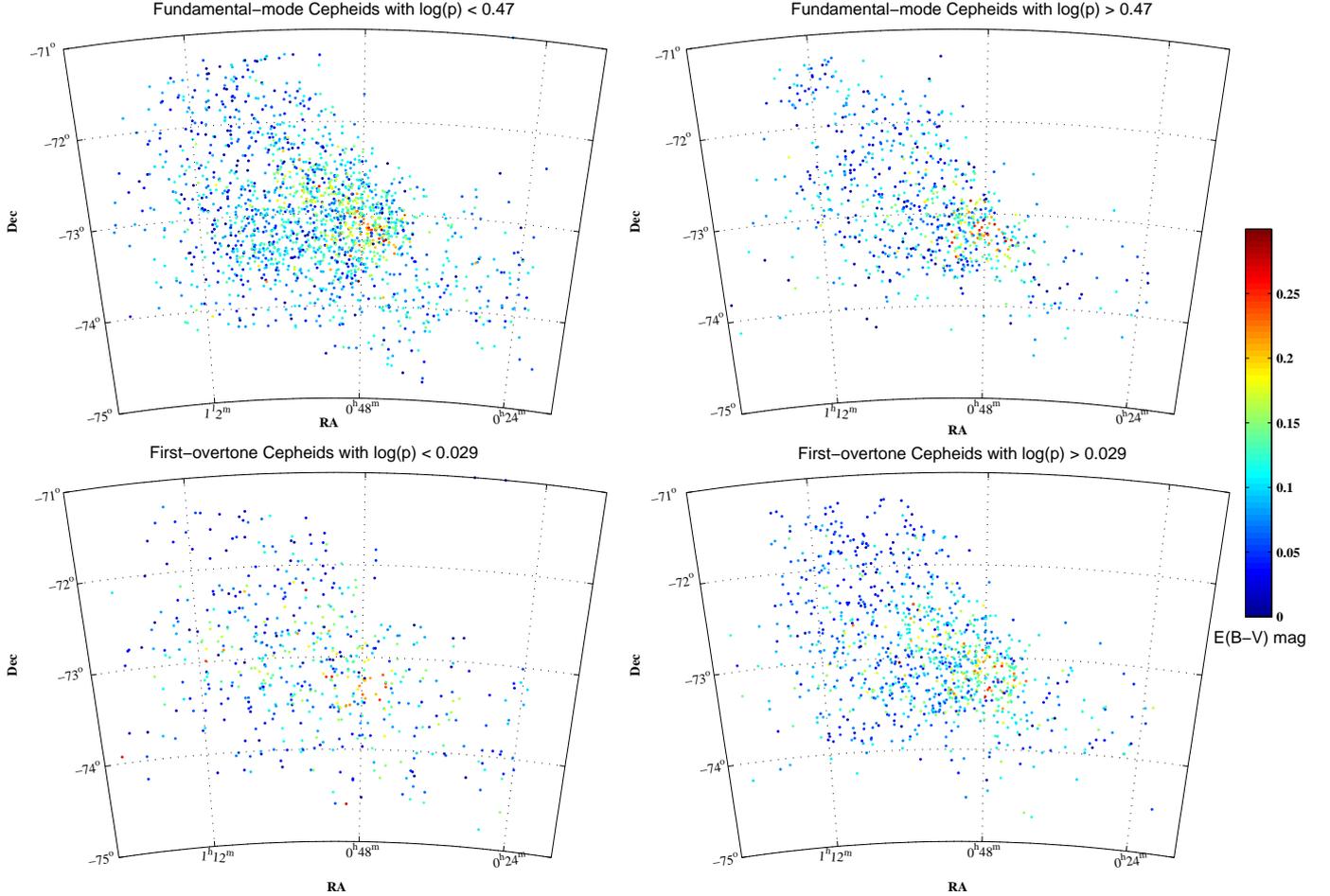}}
\centering \caption{E(B-V) reddening maps towards the SMC derived from the fundamental-mode and first-overtone Cepheids.}
\end{figure*}

\subsection{Reddening map}
The foreground  
reddening towards each Cepheid in the fundamental-mode sample and first-overtone sample 
was estimated using the relation, E$(B-V)$ = E$(B-V)_{mean}$ 
+ $\delta$E(B-V). The mean value of reddening towards the SMC was taken as E$(B-V)_{mean}$ = 0.09 \citep{Massey95}. 
The E$(B-V)$ estimated from the whole sample consisting of fundamental-mode and first-over tone Cepheids ranges from 
0.03 to 0.27 mag with 
an average value of 0.096$\pm$0.08 mag. 
The average of the reddening values estimated from fundamental-mode shorter and longer periods Cepheids separately 
are 0.096 $\pm$ 0.07 mag and 0.097$\pm$ 0.08. 
The upper left and upper right panels in Fig. 3 show the 
two-dimensional reddening maps obtained from the shorter and longer period fundamental-mode Cepheids.
The average of the reddening values estimated from 
first-overtone shorter and longer periods Cepheids separately 
are 0.087 $\pm$ 0.07 mag and 0.094$\pm$ 0.07. 
The lower left and lower right panels in Fig. 3 show the 
two-dimensional reddening maps obtained from the shorter and longer period first-overtone Cepheids.
From all the panels of Fig. 3, we can see that the central regions have a higher reddening than the surrounding 
regions. This result is consistent with the reddening map \citep{is11}  
obtained from the stars younger than 100 Myr. The reddening map obtained from the red clump stars given in 
Fig. 4 of \cite{ss12} 
also shows large reddening in the central regions. 
The mean E($B-V)$ obtained from the study of 
young stars \citep{is11} is $\sim$ 0.13 mag and from the study of 
red clump stars \cite{ss12} 
it is 0.037 mag. 
The relative variation of reddening in the SMC is 
similar in all the reddening maps obtained from different tracers. 
Even though the regions with 
high reddening coincide in the reddening maps obtained using different 
tracers, the values are different. The mean value obtained from our study is 0.096$\pm$ 0.08 mag. 
This is similar to the value obtained from the study of stellar population of similar age as that of Cepheids 
\citep{is11}. 

\begin{table*}
\centering
 \caption {Orientation measurements of the SMC disk}
\begin{tabular}{|l|l|l|}
\hline
Data & Parameters before excluding the outliers & Parameters after excluding those \\
& & which show deviation $>$ 6 kpc\\
\hline
& 
\begin{tabular}{lll}
 Number & $\it{i}$ (degrees) & $\phi$ (degrees)\\
\end{tabular}
&
\begin{tabular}{lll}
 Number & $\it{i}$ (degrees) & $\phi$ (degrees)\\
\end{tabular}
\\
\hline

\bf{Fundamental-mode Cepheids} & & \\
\hline

Cepheids with log(P) $<$ 0.47 &
\begin{tabular}{lll}
 1764 & 63.1 $\pm$ 0.98  & 156.1 $\pm$ 8.8 \\
\end{tabular}
&
\begin{tabular}{lll}
 1549 & 63.2 $\pm$ 1.09 & 151.4 $\pm$ 9.2 \\
\end{tabular}
\\

Cepheids with log(P) $>$ 0.47 &
\begin{tabular}{lll}
$ $ $ $ 778 & 64.9 $\pm$ 1.61 & 148.7 $\pm$ 13.7 \\
\end{tabular}
&
\begin{tabular}{lll}
$ $  $ $ 683 & 64.9 $\pm$ 1.66 & 154.9 $\pm$ 15.6 \\
\end{tabular}
\\

Combined sample &
\begin{tabular}{lll}
 2542 & 63.3 $\pm$ 0.85  & 154.5 $\pm$ 7.4 \\
\end{tabular}
&
\begin{tabular}{lll}
 2230 & 63.1 $\pm$ 0.96 & 151.5 $\pm$ 7.8 \\
\end{tabular}
\\

\hline

\bf{First-overtone Cepheids} & & \\
\hline

Cepheids with log(P) $<$ 0.029 &
\begin{tabular}{lll}
 $ $ $ $ 559 & 67.2 $\pm$ 0.90  & 174.1 $\pm$ 16.1 \\
\end{tabular}
&
\begin{tabular}{lll}
 $ $ $ $ 440 & 65.3 $\pm$ 1.21 & 174.2 $\pm$ 18.4 \\
\end{tabular}
\\

Cepheids with log(P) $>$ 0.029 &
\begin{tabular}{lll}
1041 & 66.8 $\pm$ 1.04 & 154.3 $\pm$ 11.5 \\
\end{tabular}
&
\begin{tabular}{lll}
$ $ $ $ 927 & 68.4 $\pm$ 0.95 & 156.3 $\pm$ 12.4 \\
\end{tabular}
\\

Combined sample &
\begin{tabular}{lll}
 1600 & 66.7 $\pm$ 0.72  & 163.0 $\pm$ 9.6 \\
\end{tabular}
&
\begin{tabular}{lll}
 1367 & 66.3 $\pm$ 0.86 & 159.3 $\pm$ 10.2 \\
\end{tabular}
\\

\hline

\bf {Combined sample of fundamental-mode and first-overtone} 
&
\begin{tabular}{lll}
 4142 & 64.6 $\pm$ 0.58  & 158.3 $\pm$ 5.9 \\
\end{tabular}
&
\begin{tabular}{lll}
 3591 & 64.4 $\pm$ 0.66 & 155.3 $\pm$ 6.3 \\
\end{tabular}
\\

\hline

\end{tabular}

\end{table*}
\subsection{Orientation measurements}
The SMC disk was modelled to have a two dimensional planar geometry. Disk parameters 
such as the inclination $\it{i}$ and the position angle of the line of nodes, $\phi$, can 
be estimated by solving the equation of the plane of the disk, as explained in Sect. 3. 
The Cartesian coordinates (xyz) of the Cepheids in our sample were estimated in the previous section. 
A plane-fitting procedure was applied to these Cartesian 
coordinates. From the coefficients of the equation of the plane, the structural 
parameters were estimated. These parameters were 
estimated separately using the longer and shorter period sample of fundamental-mode and first-overtone 
Cepheids, the combined sample 
of fundamental-mode Cepheids, the combined sample of first-overtone Cepheids, and 
finally for the full sample.  
The extra-planar features in the SMC disk were estimated as explained in section 3. 
The average error in z is $\sim$ 2.8 kpc and the deviations greater than 6 kpc (which 
correspond to more than twice the error in z) were considered as real deviations. 
After removing outliers from the sample, the planar parameters for the sub-groups were re-estimated. 
The parameters obtained are given in Table 3. From the table we can see that the 
$\phi$ values and ${\it i}$ values have a range, but they match within errors. 
\begin{figure*}
\resizebox{\hsize}{!}{\includegraphics{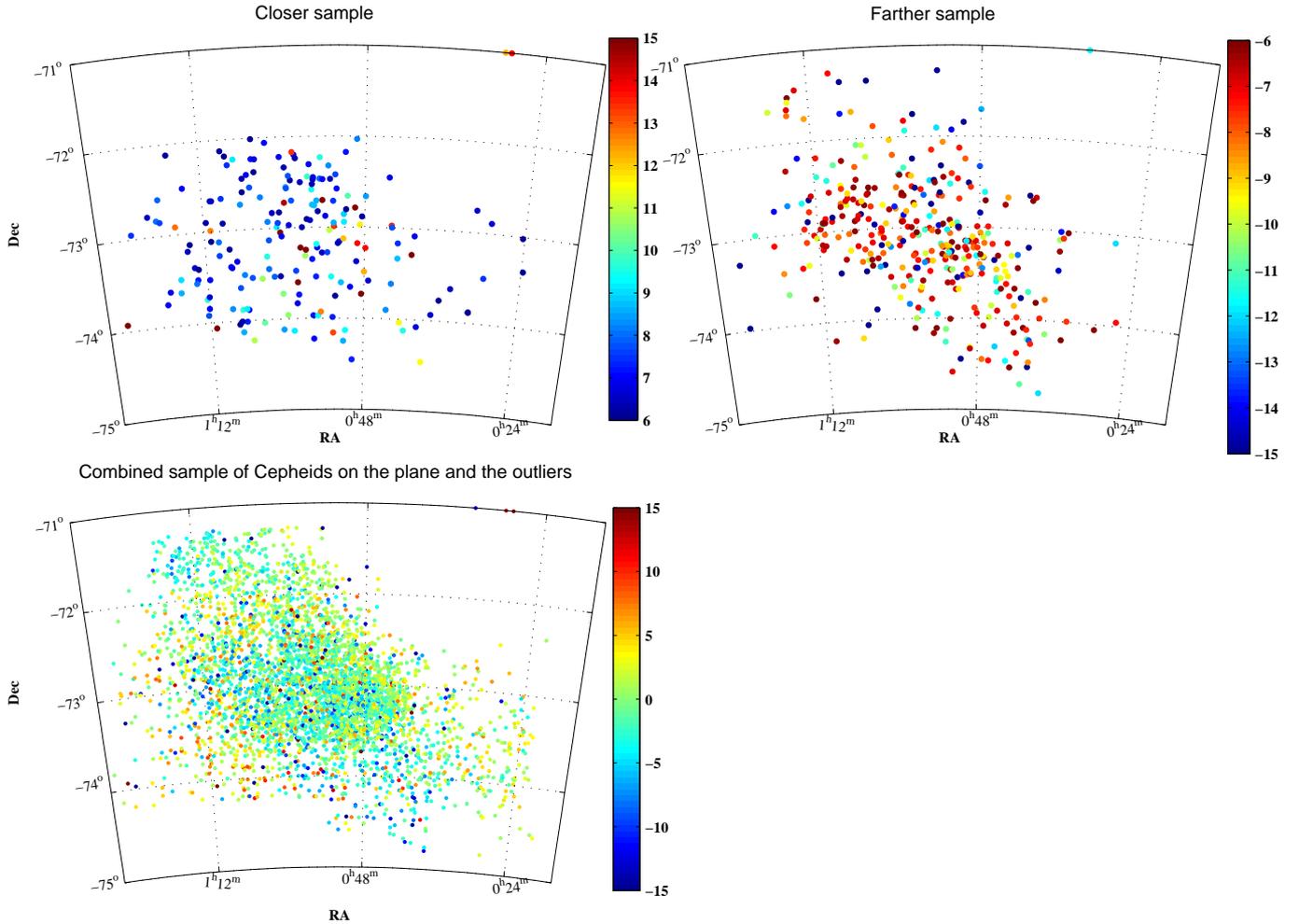}}
\caption{Two-dimensional plot of deviation. The colour bars in all the panels represent the values of the 
deviation in kpc from the best-fit plane to the SMC disk. The sample that shows absolute deviation larger 
than 6 kpc are considered as outliers from the plane, which implies that the sample with deviations in the 
range -6 kpc to +6 kpc is on the plane. The closer sample is that which shows deviation larger than
+6 kpc and the farther sample is that which shows deviation less than -6 kpc. 
 }
\end{figure*}

To understand the locations where Cepheids deviate strongly from the 
plane, we plotted the two-dimensional plots of the deviations obtained for the entire sample in Fig. 4. 
Of 4142 Cepheids, 551 (13.3\%) are out of the plane. 
There are 192 Cepheids (5\% of the total sample and 35\% of the outliers) 
in front of the fitted plane. There are 359 Cepheids (9\% of the total 
sample and 65\% of the outliers) behind the fitted plane. Comparing these two plots, we can see that the number 
of closer and farther sample in the central region is almost similar, whereas the northern and southern regions have 
more farther Cepheids and the eastern and western regions have more closer Cepheids.  The extra planar features are later 
discussed in detail in Sect. 5.3.  

The radial variation of the disk parameters up to $\sim$ 3 degrees of the SMC disk is studied using the entire 
(fundamental-mode and first-overtone) sample and is shown in Fig. 5. 
The radial variation of inclination shows that in the inner disk 
(r $< \sim$2.5 degrees)
$\it{i}$ gradually decreases and then becomes almost constant in the outer disk. On the other hand, 
$\phi$ in the inner disk (r $ < \sim$2.5 degrees) gradually increases and then becomes almost constant 
in the outer disk. 
There is a mild indication of structures/disturbances in the inner SMC (0.5 $<$ r $<$ 2.5 degree). Beyond 
r $>$ 2.5, the SMC disk seems to be stable and less structured/disturbed. 
A similar study of the outer disk of the SMC, which is beyond the scope of 
this study (r $>$ 3 degree), will be useful to understand the structural variations in the outer disk. 

The z values of the combined sample of fundamental-mode and first-overtone Cepheids are 
plotted along the axis perpendicular 
to the line of nodes, the axis of steepest gradient and is shown in Fig. 6. 
The effect 
of inclination is clearly seen in the plot. The slope estimated 
for the gradient is 1.91$\pm$0.05, which converts into an inclination of 
62$^\circ$.4 $\pm$ 2$^\circ$.9. The inclination obtained matches the inclination 
values obtained from the plane-fitting procedure within errors.\\

\begin{figure}
\resizebox{\hsize}{!}{\includegraphics{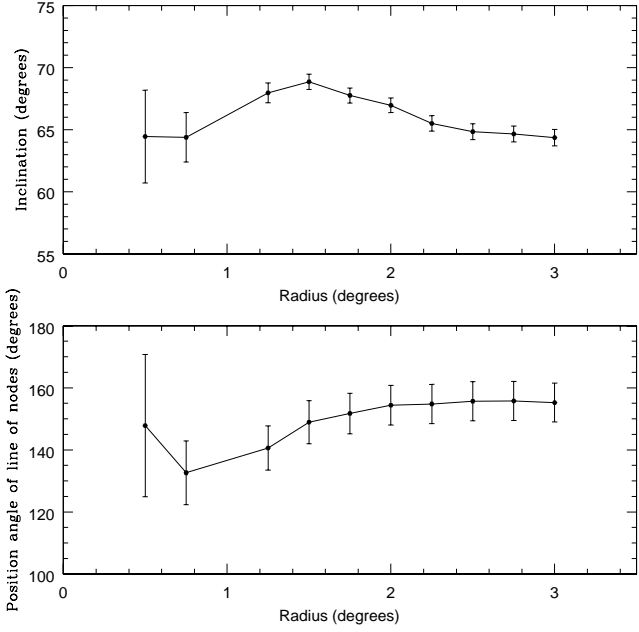}}
\caption{Radial variation of the inclination and position angle of the line of nodes 
of the SMC disk. }
\end{figure}

\begin{figure}
\resizebox{\hsize}{!}{\includegraphics{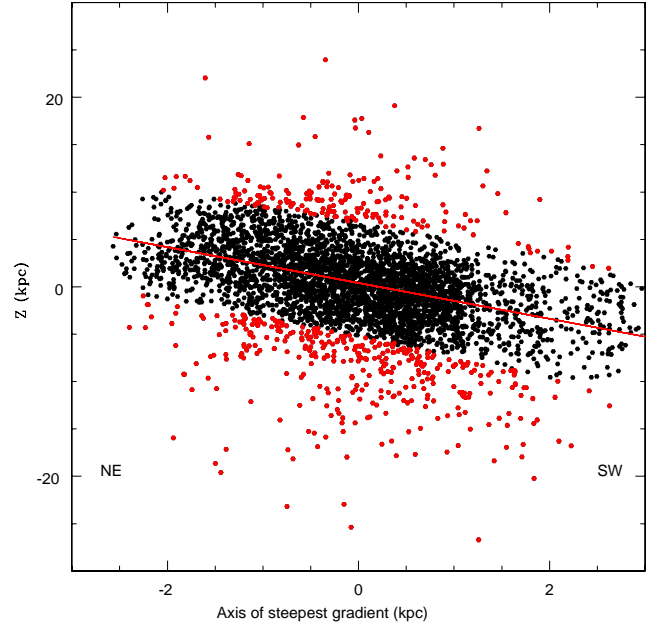}}
\caption{Relative distances (z) are plotted against the axis of steepest gradient. 
The red points are stars which show deviation higher than 6 kpc. The 
direction of inclination is shown as a red line. }
\end{figure}

After obtaining the orientation measurements of the SMC disk with respect to the sky plane, we can use the 
transformation equations 
to derive the (x',y',z') coordinates in the SMC plane. The structural parameters 
obtained from the combined sample of fundamental-mode and first-overtone Cepheids were used for the transformation. 
The x'y'z' system was obtained by two rotations of the xyz system, first a counter-clockwise rotation 
around the z axis by an angle ($\phi$ + 90) then by a clockwise rotation around the new x' axis 
by an angle $\it{i}$. The x' axis is along the line of nodes of the SMC disk and the y' axis is perpendicular 
to x'. The x'-y' plane is the actual SMC plane and is defined by the perpendicular axis, z' = 0. The range of 
z' values obtained using transformation equations is the actual measure of the intrinsic depth/thickness of the disk. 
The standard convention of z' is such that the positive and negative values represent the direction in front and behind the 
SMC plane. 
\begin{figure}
\resizebox{\hsize}{!}{\includegraphics{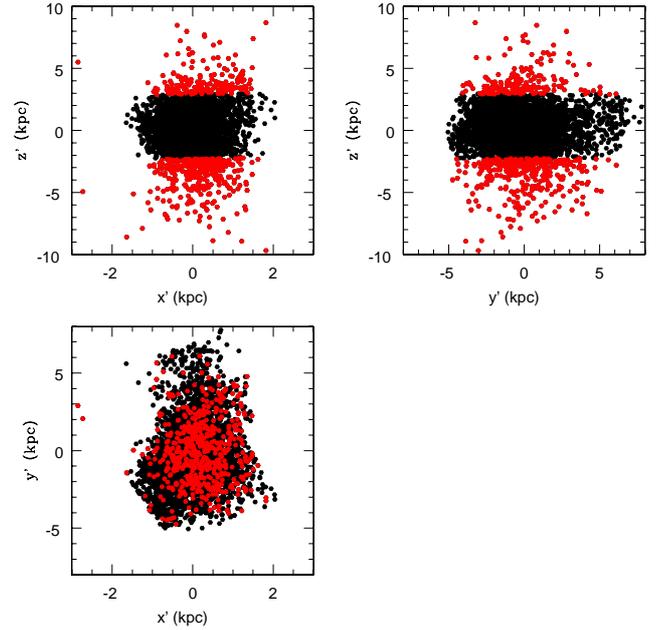}}
\caption{x'-y', x'-z' and y'-z' distributions of the SMC Cepheids are shown in the left lower, left upper, 
and right upper panels, respectively. In all the panels, the black points are the sample on the fitted plane 
and the red points are the outliers.}
\end{figure}

The relative distance (z) of the Cepheids in our sample with respect to the mean distance to the SMC gives 
an estimate of the LOS depth of the disk. But this value will over-estimate the depth/thickness 
of the disk because of its orientation with respect to the sky plane, especially its high inclination. 
The depth corrected for orientation effects is the intrinsic depth/thickness of the disk. 
As mentioned in the previous paragraph, the range of z' values obtained using transformation equations 
is the actual measure of the intrinsic depth/thickness of the disk. 
Figure 8 shows the z' and z distributions 
of the SMC disk. The 2$\sigma_{z}$ and 2$\sigma_{z'}$ values of the z and z'
distributions can be taken as the measure of the LOS depth and the orientation-corrected depth/thickness of 
the disk. These have to corrected for the 
error associated with the estimate of z and z'. The average error,$\sigma_{dis}$
 associated with the estimate of z and z' is $\sim$ 2.8 kpc. Thus the \\\\
LOS depth = $((2\sigma_{z})^2 - \sigma_{dis}^2)^{0.5}$.\\\\
The orientation-corrected depth/thickness = $((2\sigma_{z'})^2 - \sigma_{dis}^2)^{0.5}$.\\\\
The LOS depth estimated based on the z distribution is 8.1 $\pm$ 1.4 kpc. H12 estimated 
a LOS depth of 7.5 $\pm$ 0.3 kpc, based on the study of fundamental-mode Cepheids in the OGLE III field which 
matches our estimates within errors. The orientation-corrected depth/thickness estimated 
based on z' distribution is 1.76 $\pm$ 0.6 kpc. The scale height of the SMC disk can be calculated from 
the depth estimates using the relation scale height = 0.4648 * depth, given in H12. 
Thus the scale height of the SMC disk is found to be 0.82 $\pm$ 0.3 kpc.

The scale height of the SMC disk can be compared with that of the LMC and the Milky Way. 
From studying the Cepheids, \cite{has12b} estimated the scale height of the LMC disk 
as 0.8$\pm$0.2 kpc. The 
scale height of the young population in the thin disk of our Galaxy is $\sim$ 100 pc. 
The values indicate that the scale height decreases as the mass of the system increases. This is consistent with the  
results of \cite{Seth05}, who reported that lower mass galaxies form stars in a thicker disk than high mass galaxies. 
The mutual interactions with the MCs could have made the disks of these galaxies thicker. 
%
\begin{figure}
\resizebox{\hsize}{!}{\includegraphics{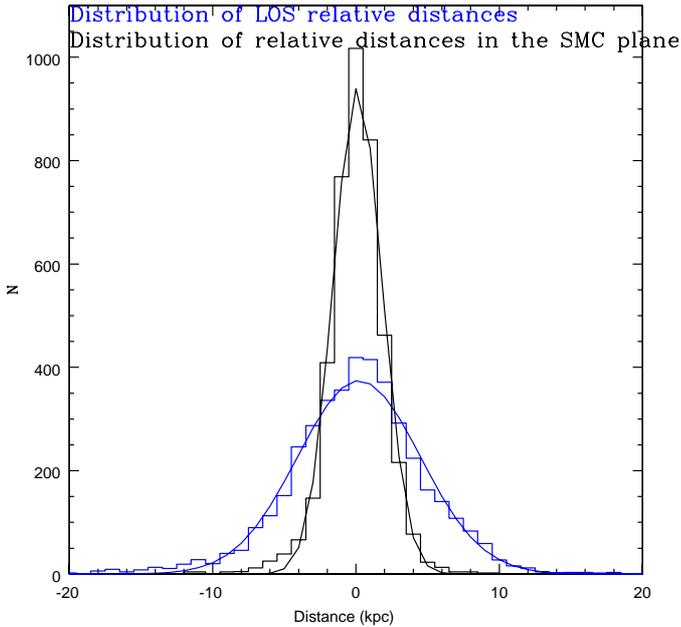}}
\caption{Number distributions of the z and z' estimates of Cepheids. Black and blue represent 
the z' and z distributions. The blue and black lines represent the best Gaussian fits 
to the respective distributions.}
\end{figure}

\subsection{Age distribution of Cepheids}
The ages of the fundamental-mode and fisrt-over tone Cepheids in the sample were estimated using the 
respective PAC relation given in Sect. 3.2. The sample stars have an age range of 10-900 Myr. The average 
error in age is $\sim$ 20 Myr. 
The age distribution of the total sample is shown in Fig. 9. 
We restricted the sample to the age of 500 Myr as very few Cepheids are older than 500 Myr.  
\begin{figure}
\resizebox{\hsize}{!}{\includegraphics{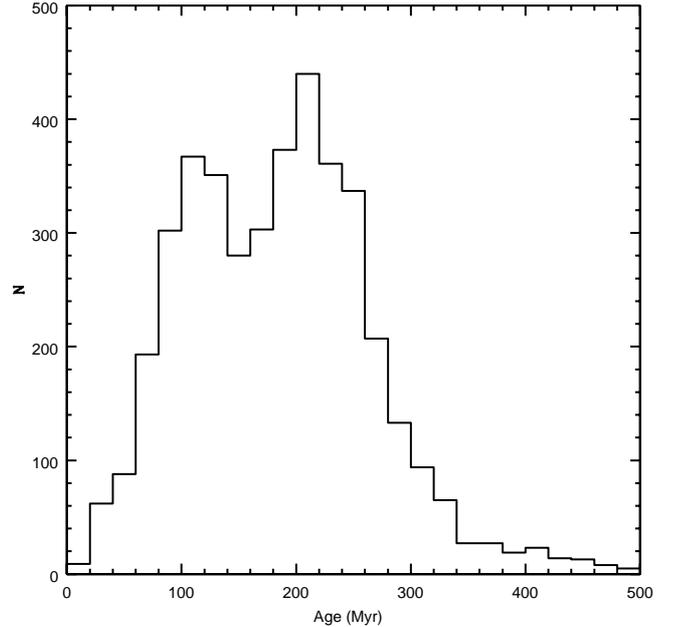}}
\caption{Age distribution of Cepheids in the SMC.}
\end{figure}
\begin{figure*}
\resizebox{\hsize}{!}{\includegraphics{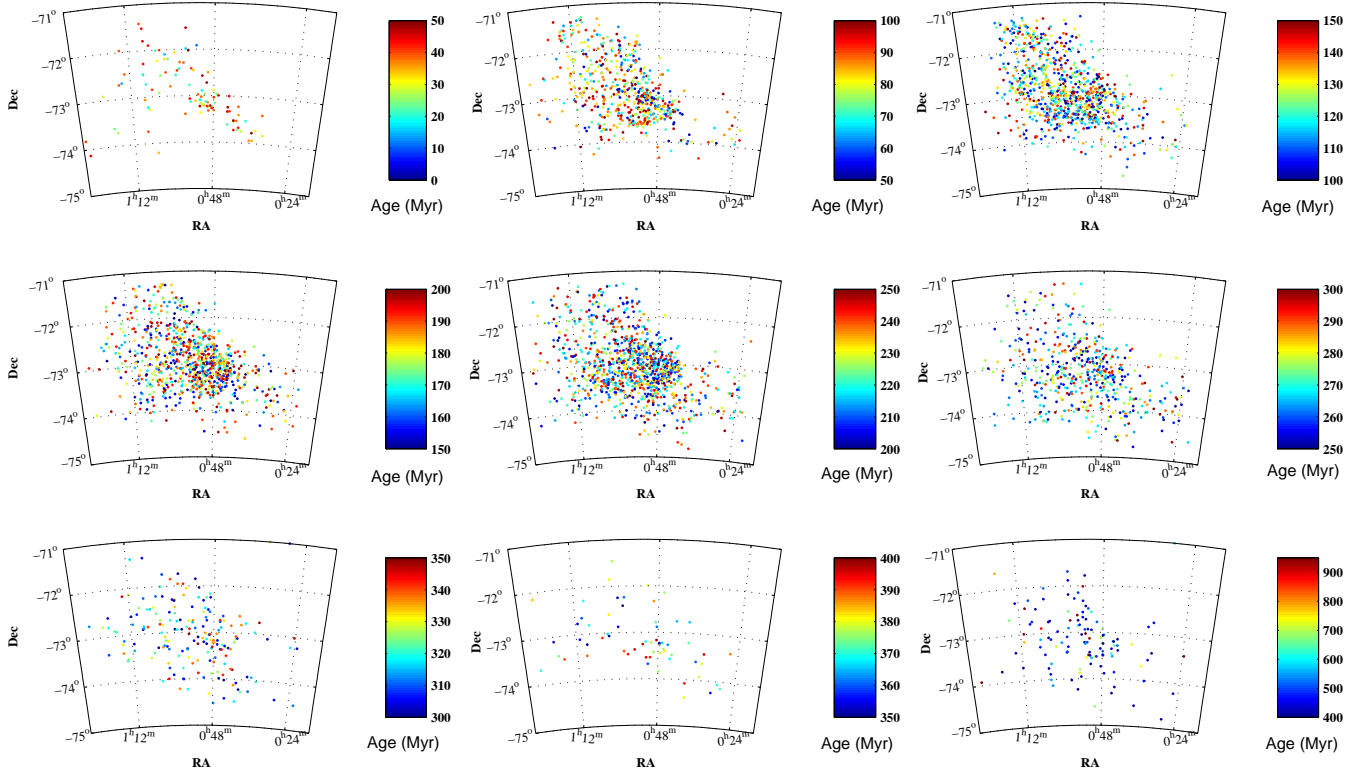}}
\caption{Two-dimensional age map of the Cepheids in the SMC (derived based on the PAC relations of the  
fundamental-mode and first-overtone Cepheids). The stars are shown in 9 different age bins. The age bin size in  
all the panels except the one in bottom right panel is 50 Myr. The bottom right panel shows stars with age $>$ 400 Myr.
}
\end{figure*}
The age distribution in Fig. 9 shows two peaks, the main peak at 200-220 Myr and another smaller peak at 
100-140 Myr. 
\cite{pu00} found two peaks in the age distribution of SMC young clusters, 
at $\sim$ 100 Myr and $\sim$ 160 Myr. 
\cite{gla10} found a peak of cluster formation around 160 Myr in the SMC. 
We inspected the age distribution plots (Fig. 1 of \citealt{pu00} and Fig. 5 of \citealt{gla10}) obtained in 
the previous studies. Figure 1 of 
\cite{pu00} and the lower panel of Fig.5 of \cite{gla10} show that there are two 
peaks in the age range 100-200 Myr in the SMC with a dip in between. Thus the general 
profile of the age distribution of Cepheids matches the age distribution of clusters.  


Two-dimensional plots of the age distribution of Cepheids in the SMC are shown in Fig. 10. 
The age bin size in all the panels except the right bottom panel is 50 Myr. In the right bottom panel, 
the sample with age $>$ 400 Myr are shown. These older Cepheids are mainly from the shorter period first-overtone 
sample. The figure shows the spatial distribution of Cepheids 
at different epochs. 
From the upper middle panel it can be seen that the Cepheids with age$\le$ 100 Myr are concentrated in the 
northern and eastern regions compared with the southern and western regions. 
The upper left panel shows that the younger Cepheids, age$\le$50 Myr are confined to the central bar region.
\section{Discussion}
\subsection{Break in PL relations}
We estimated a break in the PL relation of 
fundamental-mode Cepheids and first-overtone Cepheids at $\sim$ 2.95 days (log(P) = 0.47) and  at $\sim$ 1 day 
(log(P) = 0.0293) by applying two linear regression fits. Because the break in PL relation is associated  
with the break in period-colour relation \citep{nk06}, we investigated the break in period colour 
relation. Fig. 11 shows a two linear fit regression analysis for the period - (V-I) colour. 
The break points estimated are more or less similar to the break points estimated in PL relations. The 
break in period-colour relation at similar points confirms the break in PL relations. 
 
\begin{figure}
\resizebox{\hsize}{!}{\includegraphics{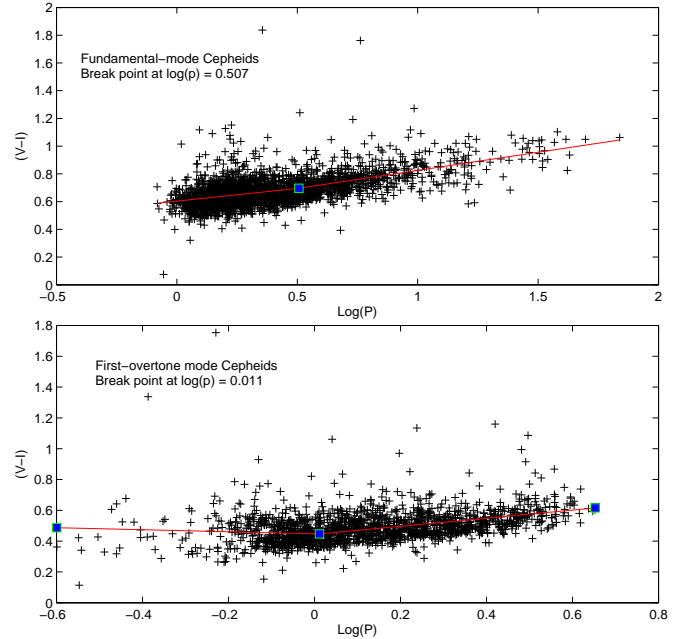}}
\caption{$(V-I)$ period-colour diagrams of fundamental-mode and first-overtone Cepheids. 
The estimated break points are also shown as blue squares.}
\end{figure}

\begin{figure}
\resizebox{\hsize}{!}{\includegraphics{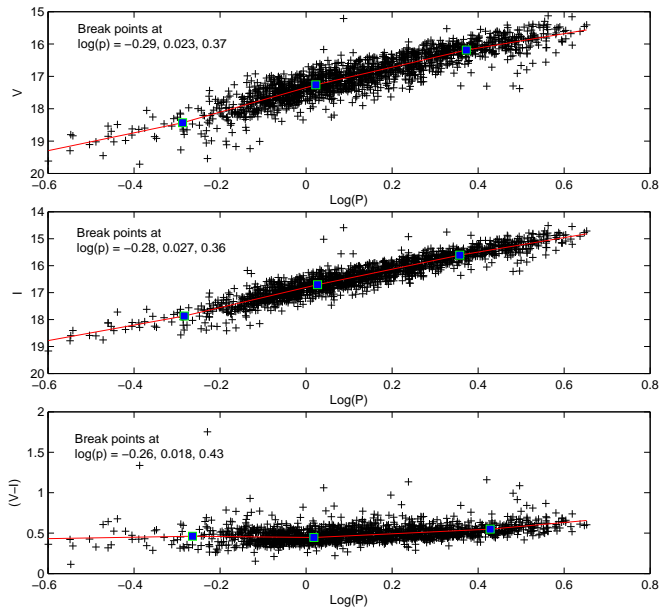}}
\caption{PL relation in VI bands and $(V-I)$ period-colour diagrams of 
first-overtone Cepheids. The multiple linear regression fits are shown as red lines, 
the estimated break points as blue squares.}
\end{figure}

As mentioned in Sect. 4.1, \cite{tam11} performed two linear regression analyses of the PL 
and period-colour relations of Cepheids in the bar region of the SMC. The break point estimated by 
\cite{tam11} for the fundamental-mode Cepheids is close to what 
we have obtained. But the break estimated for the first-overtone by them differs from that of our estimates. 
The range of log(P) values of fundamental-mode Cepheids in both the studies are same (0 $<$ log(p) $<$ 1.6). 
But for the first-overtone sample, our study has a larger sample in the shorter period range (log(P) $<$ 0.0).
The range of log(P) values of first-overtone sample in our study is -0.4 $<$ log(P) $<$ 0.7, whereas the 
range of log(P) values of first-overtone sample in the study of \cite{tam11} is -0.2 $<$ log(P) $<$ 0.7.  
From Fig. 2 of \cite{tam11} we can see that although the prominent break 
point is at log(P) $\sim$ 0.4, there is another break at log(P) $\sim$ 0. This break point was not identified in 
their study because of the smaller sample size in the shorter period range. A larger sample 
in the shorter period range would have made the break point at log(P) $\sim$ 0.029 prominent in our study. 
To estimate the other significant break points in our sample, we performed a multiple regression analysis 
to the first-overtone sample. Fig. 12 shows the PL fits and period-colour fits. We estimated three break points at similar 
points for all the three fits. The values of the break points are given in the plot. The plot indicates that 
there are three break points in our sample at log(P) $\sim$ -0.28, 0.02, and 0.4. When multiple regression 
is applied, along with the prominent break point at log(P) = 0.02, we estimate the other significant break 
points at log(P) at -0.28, and 0.4. The break point estimated by \cite{tam11} at log(P) = 0.4 
is obtained in our sample when a multiple regression analysis is performed. The difference in the sample size might 
be the reason for the differences in the estimates of prominent break points in the two studies. The sample of 
\cite{tam11} covered the bar region of the SMC, whereas our sample covers the bar and the disk 
region. The prominent break points separates the two major different populations in the region. Thus we can say 
that the major populations in the bar and disk regions are different. The disk region has more shorter period (log(P) $<$ 0) 
first-overtone Cepheids that are significantly different from the longer period ones. 

The ages corresponding to the break points in the PL relations of fundamental-mode and 
first-overtone Cepheids were estimated using PAC relation to be $\sim$ 125 Myr and $\sim$ 250 Myr. 
The age can be associated with the mass of the star, which means that the evolution of a star with different 
mass and composition can contribute to the observed break in PL relation. Fig. 1(b) of \cite{cor03} shows the 
location of OGLE II identified SMC Cepheids in the colour-magnitude diagram along with the Z=0.004 metallicity mass 
tracks. They found that the blue loops of three solar mass evolutionary tracks of SMC metallicity do not corss the 
instability strip (IS) and hence stars fainter than I $\sim$ 16 mag cannot form Cepheids with SMC metallicity. 
On the other hand, the OGLE II sample and our sample have many Cepheids faniter than 
I$=$16.5, up to about I$=$18.0 mag. \cite{cor03} claimed that as the evolutionaty tracks 
of lesser metallicty (Z=0.001) cross the IS for fainter magnitudes, the fainter Cepheids are all 
of lower metallicity. They mention that among the low-luminosity stars those that are metal poor become Cepheids 
in this range. The study by \cite{va09} 
found similar results, where only the metal-poor model (Z=0.0035)  could reproduce 
the LMC Cepheids up to a period of 1.5 days. Because our data include Cepheids up to
0.3 day, Cepheids with periods shorter than 1.4 days still have to be metal poor.
We found that low-luminosity Cepheids have a slightly different P-L relation. 
The break points in the PL relation in the fundamental mode and first overtone are found at about I=16 mag 
and I=16.8 mag. This magnitude range more or less coincides with the location in which 
\cite{cor03} expected metal-poor Cepheids. Thus, it is possible that the break in the PL 
relation observed here is due to metal-poor nature of the fainter Cepheids. 
The nature of the slope for the fainter ones also suggest that, with respect to the brighter Cepheids, the fainter
ones are intrinsically brighter, thus metal poor. Thus, the break in the PL relation might be the observational evidence 
that the fainter Cepheids in the SMC are metal poor. If this is true, 
this result also implies a very metal poor and young population in the SMC, 
along with those of SMC metallicity. Detailed high-resolution spectroscopy of these faint Cepheids is necessary to confirm this claim.

\subsection{Orientation measurements: Comparison with previous studies}
The combined sample of the fundamental-mode and first-overtone Cepheids in 
the SMC, are used to estimate the orientation measurements of the SMC disk. 
From Table 3 we can see that the analysis of different sub-groups (shorter and longer fundamental-mode 
Cepheids and shorter and longer first-overtone Cepheids) in our sample gives similar 
estimates for the orientation measurements of the SMC disk. This suggests that there 
is no significant change in the disk structure of the SMC as traced by different sub-groups in our sample.
The results obtained here (for the combined sample) is compared with previous 
estimates in Table 4. \cite{g00} have used a different definition/notation for 
the position angle of line of nodes, which makes their position angle value 90$^\circ$ 
rotated in the counter clockwise direction. When this is considered, our results are consistent 
with previous estimates derived from Cepheids. All the estimates from 
Cepheids (other than that of H12) are based on studies of the central region of the SMC, 
mainly the bar region. Compared with these previous studies, our sample has more Cepheids. 

\cite{stan04}, based on H {\sc i}, reported a $\phi$ of $\sim$ 45$^\circ$ for the 
H {\sc i} disk of the SMC. The $\phi$ obtained in our analysis is almost orthogonal to that 
seen in the H {\sc i}. This suggests that the kinematical line of nodes is 
perpendicular to the photometric line of nodes. \cite{eh08}  
obtained a position angle  ($\sim$ 126$^o$)  for the line of 
steepest velocity gradient, which is similar to the $\phi$ we obtained.  
A recent kinematic study of red giants and F/G super giants by \cite{do14} found 
$\phi$ to be $\sim$ 120$^\circ$-130$^\circ$, which is also similar to our estimates. 
A study based on a sample from a larger spatial coverage data set 
is required to understand the discrepancy in the parameters of the gaseous and stellar disks of the SMC. 

\begin{table*}
\centering
\caption{Summary of orientation measurements of the SMC disk}
\label{Table:1}
\vspace{0.25cm}
\begin{tabular}{lrrrr}
\hline \\
Reference & Inclination, $i$  & PA$_{lon}$, $\phi$  
& Tracer used for the estimate\\ \\
\hline
\hline \\

\cite{cc86} & 70$^o$$\pm$3$^o$ & 148$^o$$\pm$10 & Cepheids\\
\cite{ls86} & 45$^o$.0$\pm$7$^o$ & 145$^o$$\pm$17$^o$ & Cepheids\\
\cite{g00} & 68$^o$.0$\pm$2$^o$.0 & 238$^o$.0 $\pm$7$^o$.0 & Cepheids\\ 
H12 & 74$^o$$\pm$9$^o$ & $-$ & Cepheids\\
\cite{eh08} & $-$ & 126$^o$.0 & Kinematics of OB stars \\
\cite{stan04} & $-$ & 45$^o$ & Kinematics of H \sc i \\ 
\cite{do14} & $-$ & 120$^o$ & Kinematics of F/G supergiants\\
Our result from the combined sample& 64.4 $\pm$ 0.66 & 155.3 $\pm$ 6.3 & Cepheids\\
\hline\\
\hline
\end{tabular}
\end{table*}

\begin{figure*}
\resizebox{\hsize}{!}{\includegraphics{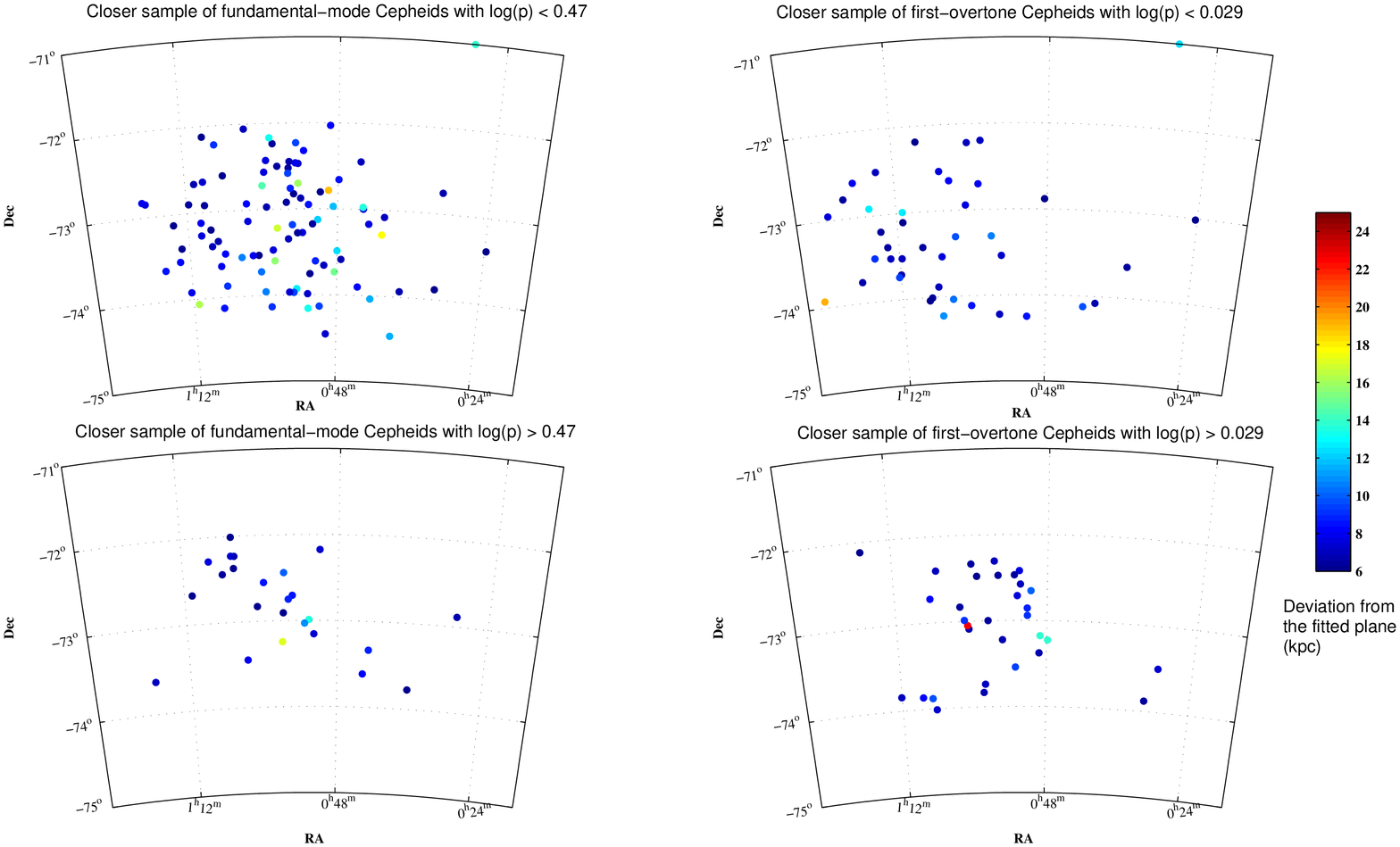}}
\caption{Two-dimensional plots of Cepheids in different sub-groups, that are in front of the fitted plane. 
The details of all the panels are given in the plot itself.}
\end{figure*}

\begin{figure*}
\resizebox{\hsize}{!}{\includegraphics{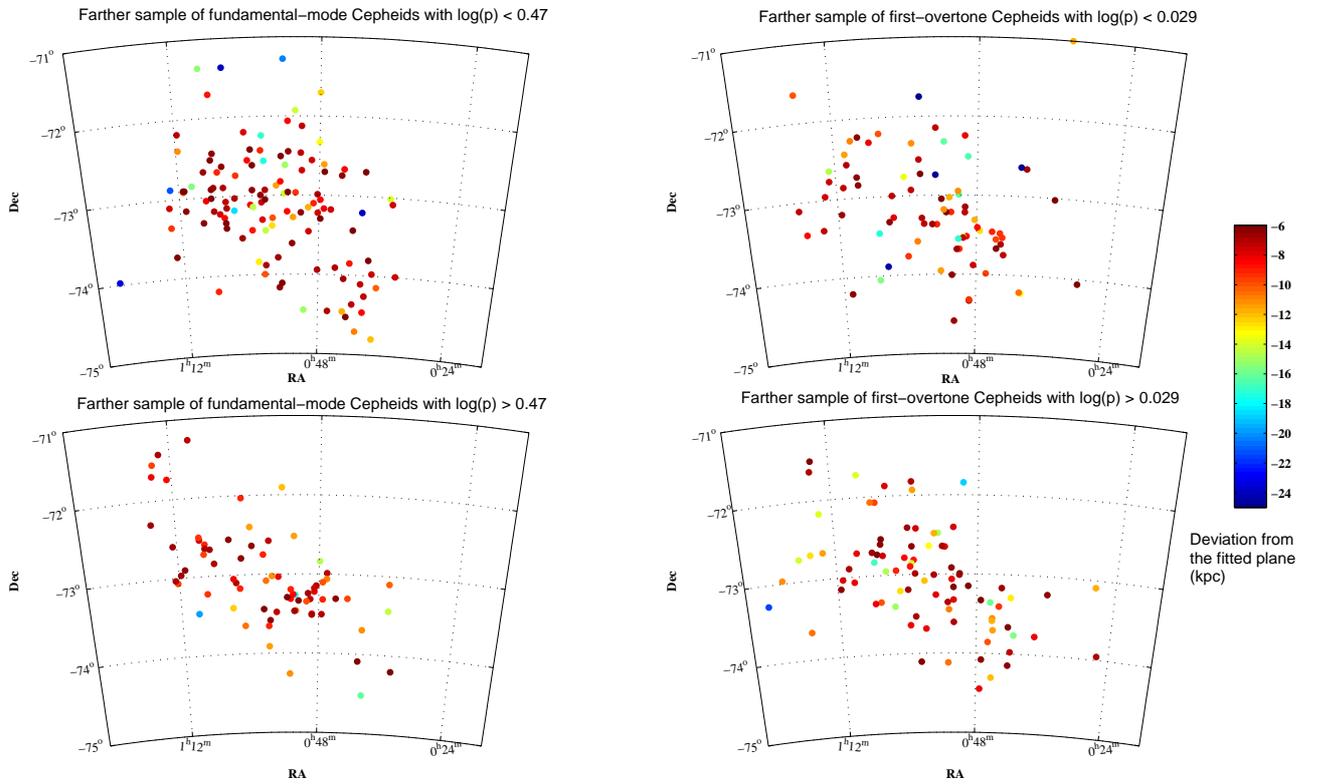}}
\caption{Two-dimensional plots of Cepheids in different sub-groups, that are behind the fitted plane. 
The details of all the panels are given in the plot 
itself.}
\end{figure*}

\subsubsection{Comparison with the study of H12}
H12 used the OGLE III data of fundamental-mode Cepheids for their study. 
Because we used the same data set in our analysis, a detailed comparison 
with their study is given below. 

H12 estimated the inclination, position angle 
of the major axis, and the LOS depth of the SMC disk. From Tables: 4 and 2 and Sect. 4.4 
we find that all the estimates from the two studies match well. 
H12 have estimated the position angle of the major axis and not the position 
angle of the line of nodes. Here we would like to 
point out that the major axis and the line of nodes are physically two different parameters.  
The major axis is the axis of elongation in the density distribution of the sample, the 
line of nodes is the line of intersection of the galaxy plane and the sky plane.
In Table 3 of H12 they have compared their value with the position angle estimates 
in the literature. The position angle values quoted in their table, from the studies of \cite{cc86} 
and \cite{ls86}, are the position angle values of closest part of the SMC disk or the position 
angle of the axis of the steepest gradient in magnitude (this axis is perpendicular to the line of nodes).   
And the position angle value from the study of \cite{g00} is the position angle of the line 
of nodes. As mentioned earlier, \cite{g00} have used a different definition for the position 
angle measurement. The values of \cite{ss12} 
quoted in Table 3 of 
H12 are the position angle of the major axis of the system. When the changes are considered, the position angle of the 
major axis of the SMC disk estimated 
by H12 matches the estimates from \cite{ss12}. But  
from the estimates of the position angle of line of nodes we can see that the major axis 
is not aligned with the line of nodes, and they are nearly orthogonal to each other. 
In the present study we also found that the major axis is perpendicular to the line of nodes. 
H12 also estimated the individual distances to each Cepheid and found that the regions 
in the eastern side of the SMC are closer to us. We also found that the eastern regions of the SMC are 
closer to us. 

In the present study there are two major additions to the study of H12. 
Along with the fundamental-mode Cepheids, we have used the first-overtone Cepheids, which increased the 
sample size. Moreover we have estimated the ages of the sample by using the PAC relation and hence estimated 
the SFH of the SMC within the last $\sim$ 50-900 Myr. Based on this we tried to understand the evolution and interaction 
history of the SMC disk. Along with the above mentioned 
major additions, there are other significant results in our present study. We estimated 
a break in the PL relations of the fundamental-mode and first-overtone Cepheids and separate PL relations 
for the sub-groups were derived and used for the estimation of relative distances. This improved the 
accuracy of structural parameters. As a by-product of the study we have presented a foreground-reddening map 
towards the SMC. The LOS depth was corrected for the effects of orientation measurements and then the scale height 
of the SMC disk was estimated.  

\subsection{Extra-planar features}
As described in Sect. 4.4, the Cepheids with a deviation larger than 6 kpc (more than 
twice the error in the estimate of the z coordinate) from the fitted plane 
were considered as extra-planar. The locations of the sample that are closer and farther from the fitted 
plane are shown in Fig. 4. The extra-planar features identified in the disk of the SMC are important because 
they give clues to the interaction and evolutionary history of the SMC disk. The effects of the population 
differences of the Cepheids and reddening in detecting the extra-planar features 
have to be considered carefully to understand the deviating structures and also to know whether the 
identified structures are real or not. In this section we discuss these effects in detail.

We obtained a break in the PL relations of the fundamental-mode and first-overtone sample of Cepheids. 
The break in the PL relation suggests that the SMC disk hosts a different population of Cepheids. 
These population differences can affect the estimate of relative distances and hence the extra-planar 
features. But in the present study, we have estimated separate PL relations for each sub-group and used 
the respective PL relations for the estimate of relative distances. Thus in principle, the population 
differences of the sub-groups should not affect the identification of extra-planar features.  
Still, it is important to understand the spatial distribution of the extra-planar features identified 
in different sub-groups. The extra-planar features, in front and behind the 
fitted plane, of different sub-groups (shorter/longer period fundamental-mode and first-overtone Cepheids) 
are shown in Figs. 13 and 14. From Fig. 13 we can see that longer period 
Cepheids in the fundamental-mode closer sample are concentrated in the 
central/bar region, with very few stars in the eastern regions. All other sub-groups are distributed 
in the central and eastern regions. From Fig. 14 we can see that the distribution of farther 
Cepheids are more or less similar for all the sub-groups. There is no particular pattern seen for 
a specific sub-group. 
\begin{figure}
\resizebox{\hsize}{!}{\includegraphics{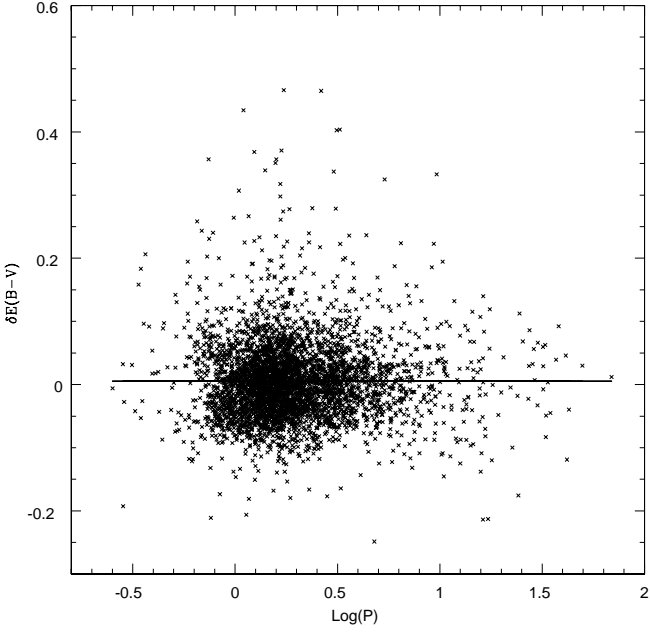}}
\caption{Relative reddeing estimates plotted against the Log(P). The best-fit line is also shown.}
\end{figure}

\begin{figure}
\resizebox{\hsize}{!}{\includegraphics{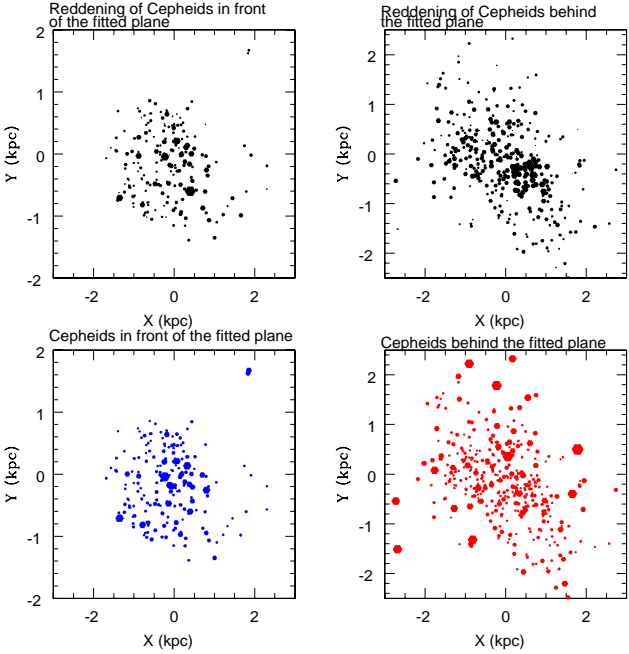}}
\caption{Lower left and lower right panels show the Cepheid distribution of in front of the plane and 
behind the plane. The size of the dots in lower panels is proportional to the deviation amplitude. 
The upper left and upper right panels show the reddening distribution. 
The size of the point in upper panels is proportional to the reddening amplitude.}
\end{figure}
 
The effect of reddening plays a main role in determining the extra-planar features. The extra-planar 
features, which are found both behind the disk and in front of the disk, could be in the 
plane of the SMC disk itself if there were an over-estimate or under-estimate of the reddening. 
In the present analysis, we have used only the PL relations, and the magnitude residuals were modelled as 
variations in distance and reddening only. Because we did not consider the period-colour relation
of the Cepheids, we could have over-estimated the reddening for long-period stars and under-estimated 
the reddening for short-period Cepheids. Fig. 15 shows the relative variation of reddening with respect to the mean 
reddening as a function of period. 
This diagram addresses the effect of any systematic effects due to the Cepheid period-colour relation 
in the derived reddenings. The fitted line in the plot shows that both the slope and intercept are zero. 
This suggests that our estimated reddening values have no systematic effect due to the period - colour relation 
of Cepheids. We also plot a two-dimensional plot of reddening and the deviations. The deviations and 
the reddening values obtained  are shown in Fig. 16. 
We do not see a strong correlation between reddening and the deviation because 
both positive and negative deviations are observed as regions with high reddening, that is the
reddening could not have been under, and over-estimated.


As this analysis show that the extra-planar features can be considered as real features, 
we carefully examined the locations of these features and tried to connect them with the already known 
features in the SMC. 
Figure 13 shows that the central and eastern regions have features in front of the plane. 
Very few stars are found to be north 
of -72$^\circ$.0. There are a few stars in the south-western region of the disk, which matches the location 
in which \cite{cc86} found some Cepheids to be in front of the disk. 
All the panels show that there are more stars in the eastern region located in fornt of the 
fitted plane by about 5-14 kpc. Recently, \cite{nid13} found a  bimodal distance distribution of the 
red clump population (age $\sim$ 1-9 Gyr) in the eastern regions (at position angles, 26$^\circ$, 
71$^\circ$, 116$^\circ$ and 161$^\circ$) of outer SMC (r = 4$^\circ$.0.), with one component in the SMC 
mean distance and the second component at 55 kpc. They suggested this component to be the tidally stripped 
stellar counterpart of H {\sc i} gas in the Magellanic Bridge. The Cepheids observed to be in front 
of the plane in the eastern regions may possibly be the younger counterpart of this tidally stripped 
population. Fig. 14 shows that the features behind the plane are mostly 
across north-east to south-west. From comparing the features in Fig. 13, we can say that 
the eastern regions have Cepheids in front of as well as behind the fitted plane. Simulations by \cite{db12} predicted a tidal 
feature known as counter bridge away from the SMC at a distance of $\sim$ 85 kpc, 
along the north-east south-west direction. Although the the Cepheids behind the plane are not at 85 kpc, 
they are possibly part of the counter bridge predicted by \cite{db12}. \cite{nid13} 
did not find intermediate-age stellar counterparts of the counter bridge. Thus the presence of Cepheids behind 
the fitted plane along the north-east to south-west is the first observational evidence for an existence 
of the counter bridge.  

The origin of young stellar population out of the plane of the SMC disk is of great interest because it 
helps to understand the recent interaction history of the SMC with its neighbours. This is discussed in 
the next section.     

\subsection{Recent epoch of interaction}
Simulations by various groups (\citealt{bes12}, \citealt{db12}) predicted the last close encounter 
between the SMC and the LMC to be around 100-300 Myr. They proposed that a direct recent collision between the 
LMC and SMC would likely leave notable marks in the SFHs of both of these galaxies. A correlated burst of star 
formation during such an  encounter has also been suggested in many earlier numerical studies 
(e.g. \citealt{gn96}; \citealt{bk05}, \citealt{bk07}). \cite{pu00} found two peaks 
in the age distribution of the LMC and the SMC young clusters, at $\sim$ 100 Myr and $\sim$ 160 Myr. 
\cite{gla10} found a peak of cluster formation around 160 Myr in the SMC and a peak at 125 Myr in 
the LMC.  A recent study of Cepheids in the LMC by \cite{jj14} found a peak of star formation around 
125 - 200 Myr, and this age range matches with our estimates of the peaks in the age distribution of Cepheids 
in the SMC. 
Thus all these observational studies found a coincidence of star formation peaks between 100-300 Myr 
in the LMC and the SMC. 

\begin{figure*}
\resizebox{\hsize}{!}{\includegraphics{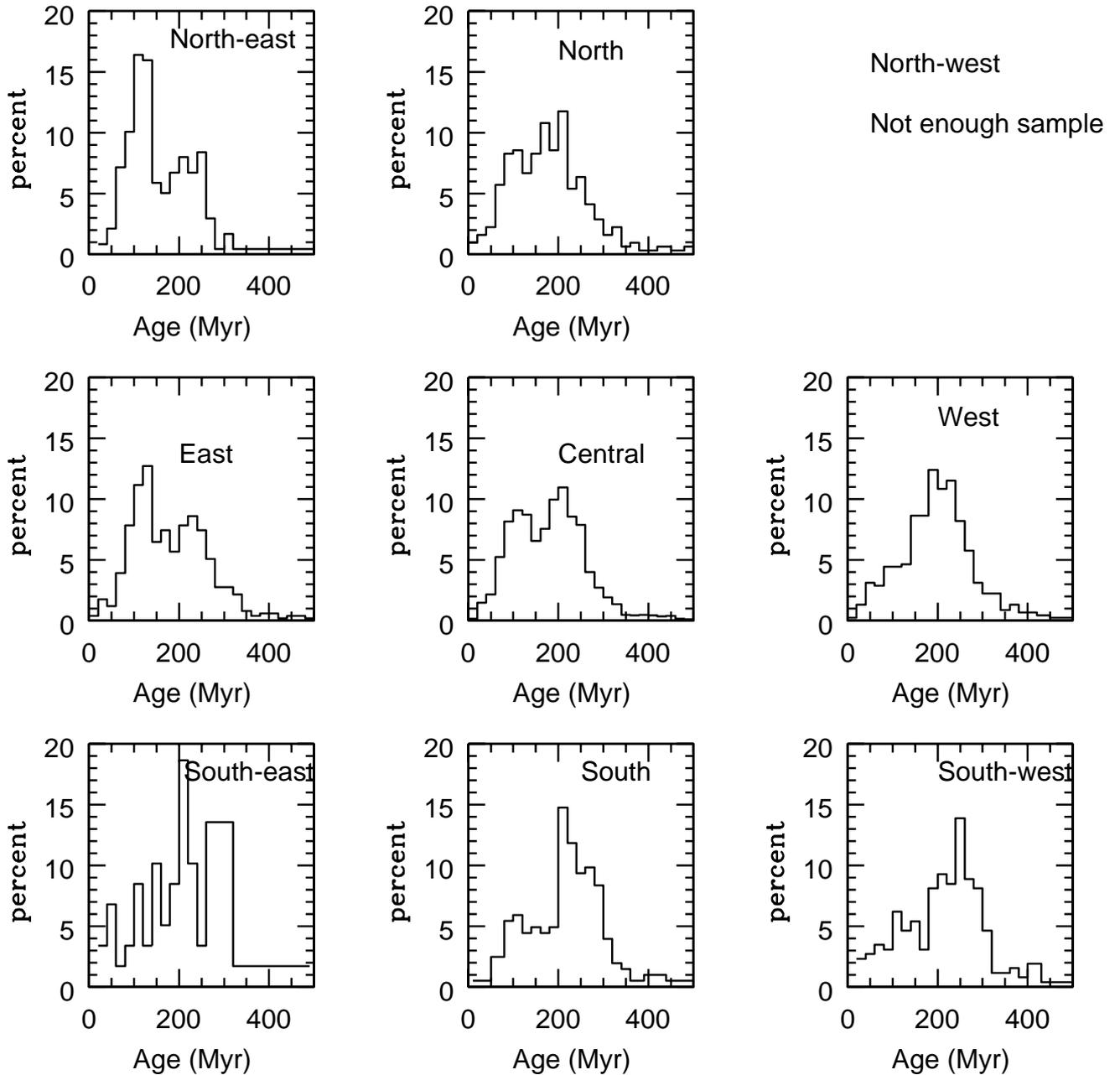}}
\caption{Normalised age distributions of Cepheids in different regions of the SMC disk.The distribution of 
the north-western region is not shown because the region does not have statistically significant sample.}
\end{figure*}

To understand the 
directional preference of star formation during the last epoch of interaction, the 
spatial dependence of age distribution of Cepheids was analysed. The 
observed region was divided into 9 regions, east (x$<$-0.75 kpc, y$<$0.75kpc and y$>$-0.75 kpc), west 
(x$>$0.75 kpc, y$<$0.75kpc and y$>$-0.75 kpc), north (y$>$0.75 kpc, x$<$0.75kpc and x$>$-0.75 kpc), 
south (y$>$-0.75 kpc, x$<$0.75kpc and x$>$-0.75 kpc), north-east (y$>$0.75 kpc and x$<$-0.75kpc), 
north-west (y$>$0.75 kpc and x$>$0.75kpc), south-east (y$<$-0.75 kpc and x$<$-0.75kpc)
south-west (y$<$-0.75 kpc and x$>$0.75kpc), and central (y$>$-0.75 kpc, y$<$0.75 kpc, x$<$0.75kpc 
and x$>$-0.75 kpc) regions. The normalised age distributions of the Cepheids in these regions 
are shown in Fig. 17. Because the number of stars in the north-western region is not statistically significant 
we do not show the age distribution of this region. Figure shows that the younger peak at $\sim$ 
100-140 Myr is not very significant 
in the western, southern, south-western, and south-eastern regions. These regions are dominated by the 
peak at earlier epoch, $\sim$ 180-260 Myr. On the other hand, the eastern and 
north-eastern regions are dominated by a peak at $\sim$ 100-140 Myr. The northern and central regions 
receive contributions from both the younger and older populations. The peak at $\sim$ 180-260 Myr 
is present in all the panels. The presence of strong younger peak in the eastern, 
especially north-eastern region mildly suggests that the star formation in the SMC might have propagated from 
south-west to north-east during the recent epoch of interaction with the LMC. \cite{is11} observed 
such a directional propagation of star formation towards the north-east in the SMC 
from analysing the young population. 

The origin of the extra-planar Cepheids that we see in our sample can be explained in the context of 
the recent LMC-SMC interaction. The recent interaction at $\sim$ 100-300 Myr ago could have triggered 
star formation in the SMC disk and could also have tidally stripped  stars and gas from the disk. 
The extra-planar Cepheids in our sample might be the tidally stripped Cepheids (which might 
have formed in the SMC disk before the epoch and/or during the early epoch of interaction). 
To determine whether the extra-planar Cepheids have any preferential age range, the normalised age distributions of the 
planar and extra-planar features were analysed. We show this in Fig. 18. The solid and dashed line show 
the age distribution of Cepheids on the fitted plane and out of the fitted plane. 
The ages of the two peaks in the distribution of the Cepheids out of the plane are similar to that identified for 
the Cepheids on the fitted plane.  
This means that the extra-planar Cepheids have no preferential age distribution and a similar 
age range as that of the Cepheids on the plane. 
The extra-planar Cepheids can be divided into three groups. One is the oldest sample with an age $>$ 300 Myr, 
the second group has an age range 100 - 300 Myr, and the third group is younger than 100 Myr. 
The classification 
is based on the criteria that the first (108 stars), second (593 stars), and third (102 stars) 
group of stars are those formed 
before, during, and after the bracketted epoch of 
recent interaction.  

We can speculate that the first group, which is the 
older population in the extra-planar sample, contains the stars tidally stripped from the plane of the SMC  
during the interaction. 
The origin of the second group of stars, which constitutes a larger portion of the extra-planar sample,  
needs to be addressed carefully. Three scenarios, associated with the epoch of interaction, which can 
be attributed to the origin of the second group, are described below. In the first scenario the 
interaction causes the gas to be tidally stripped out of the plane and stars to be formed simultaneously 
from this gas. The second scenario: interaction triggered star formation in the disk and simultaneously 
stripped a few percent of these stars out of the plane. The third scenario proposes that the interaction 
triggered star formation in the gas, which was already located out of the plane. The third scenario seems 
to be more physically feasible because the time scale involved in the first two scenarios is very short to 
explain the observed number of extra-planar Cepheids in the second group.      
Along with the stars, gas from the SMC would also have been tidally stripped 
during the recent epoch of interaction. This gas may form stars only later and not exactly at 
the epoch of interaction. Thus we can say that the third group of stars, which are the youngest 
population in the extra-planar sample, are formed from the gas tidally stripped during the recent interaction and/or 
formed from the gas that was already present out of the plane. 
Although we can argue the formation scenario of the youngest group to be from the tidally stripped gas in the 
recent or earlier interactions, the triggering mechanism involved in the formation of stars in the gas out of 
the plane is not clear. 
 
According to the model calculations by \cite{db12}, the SMC had undergone a strong interaction with the LMC 
at $\sim$ 2 Gyr ago, before the recent 
strong interaction with the LMC, which disrupted its gaseous disk. According to their model, this gas created the 
Magellanic Stream, and part of this gas also was engulfed by the LMC creating star formation in the LMC. 
The left-over gas around the SMC plane would have compressed to form stars during the recent interaction,  
which would have resulted in the observed Cepheids out of the plane  . Thus the the origin of 
the second group of extra-planar Cepheids is more likely to be due to  
the triggered star formation of the gas out of the plane around the SMC during the recent epoch of interaction. 
In addition, the presence of enough gas out of the plane as a result of SMC interaction at $\sim$ 2 Gyr ago and insitu 
star formation in the gas can explain the origin of all the three groups of extra-planar samples of Cepheids. 

If the extra-planar sample are formed from the gas out of the plane thrown out from the main disk at 
$\sim$ 2 Gyr, then these stars are expected to have a lower metallicity than the stars that formed 
in the plane at the same epoch. Thus a detailed spectroscopic study of these Cepheids 
will provide more input to understand the origin of the extra-planar Cepheids. A spectroscopic study 
will also help to distinguish the tidally stripped population from the stars that formed from the 
out of the plane gas.   

\begin{figure}
\resizebox{\hsize}{!}{\includegraphics{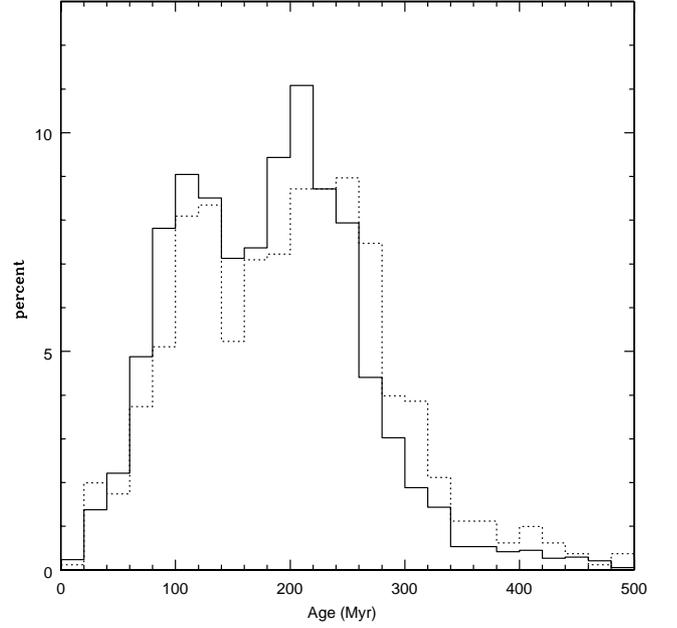}}
\caption{Normalised age distribution of the full sample of the Cepheids on the plane and 
out of the plane are shown as solid and dashed line.}
\end{figure}


\section{Summary}
We estimated the structural parameters of the SMC disk using the PL realtion 
of the fundamental-mode and first-overtone Cepheids from the OGLE III catalogue.\\ 

We estimated a break in the PL relation of fundamental mode Cepheids at log(P) = 0.47 and 
a break at log(P) = 0.029 in the PL relation of first-overtone Cepheids.\\

The planar parameters were estimated using the PL relations of shorter and longer 
period Cepheids separately. After removing the outliers, the planar parameters 
were re-estimated. For the combined sample, we found an inclination $\it{i}$ = 64$^\circ$.4$\pm$0$^\circ$.7
and $\phi$ = 155$^\circ$.3$\pm$6$^\circ$.3.\\

Extra-planar features in front and behind the fitted plane were identified. 
Some of the Cepheids in the eastern regions in front of the fitted plane are possibly 
the youngest tidally stripped population associated with the Magellanic Bridge, 
during the recent interaction of the MCs. The Cepheids behind the fitted plane provide the 
first observational evidence for an existence of the counter bridge, proposed by \cite{db12}.\\

The orientation-corrected depth/thickness of the SMC disk was estimated after correcting the 
line-of-sight depth for the orientation measurements of the disk. The scale height of the disk 
was obtained as 0.82 $\pm$ 0.3 kpc.\\

The age of Cepheids was obtained by using the PAC relation. In the age distribution we identified 
two peaks at $\sim$ 100-140 Myr and $\sim$ 200-240 Myr and the general profile of the distribution matches 
the cluster age distribution in the SMC well.\\

As a by-product of this study, a reddening map towards the SMC disk was presented.\\

The position angle $\phi$ we estimated is almost orthogonal to that obtained for the H {\sc i} disk 
of the SMC.\\

Photometric and spectroscopic studies of Cepheids in a larger area of the SMC are essential to 
better understand the variation of structural parameters and also the interaction history of the SMC.\\

\acknowledgements
The authors thank the anonymous referee for the constructive suggestions that improved the manuscript. 
The authors also thank the OGLE team for making the data available in public.


\begin{thebibliography}{}
\bibitem[Bekki \& Chiba (2005)]{bk05}Bekki, K., Chiba, M., MNRAS, 356, 680
\bibitem[Bekki \& Chiba (2007)]{bk07}Bekki, K., Chiba, M., MNRAS, 381, 16L
\bibitem[{{Bekki} \& {Chiba}(2008)}]{bc08}
{Bekki}, K. \& {Chiba}, M. 2008, \apjl, 679, L89
\bibitem[{{Bauer} {et~al.}(1999){Bauer}, {Afonso}, {Albert}, {Alard},
  {Andersen}, {Ansari}, {Aubourg}, {Bareyre}, {Beaulieu}, {Bouquet}, {Char},
  {Charlot}, {Couchot}, {Coutures}, {Derue}, {Ferlet}, {Gaucherel},
  {Glicenstein}, {Goldman}, {Gould}, {Graff}, {Gros}, {Haissinski}, {Hamilton},
  {Hardin}, {de Kat}, {Kim}, {Lasserre}, {Lesquoy}, {Loup}, {Magneville},
  {Mansoux}, {Marquette}, {Maurice}, {Milsztajn}, {Moniez},
  {Palanque-Delabrouille}, {Perdereau}, {Pr{\'e}vot}, {Renault}, {Regnault},
  {Rich}, {Spiro}, {Vidal-Madjar}, {Vigroux}, \& {Zylberajch}}]{bauer99}
{Bauer}, F., {Afonso}, C., {Albert}, J.~N., et al.  
1999, \aap, 348, 175
\bibitem[{{Besla} {et~al.}(2007){Besla}, {Kallivayalil}, {Hernquist},
  {Robertson}, {Cox}, {van der Marel}, \& {Alcock}}]{bk07}
{Besla}, G., {Kallivayalil}, N., {Hernquist}, L., {Robertson}, B., {Cox},
  T.~J., {van der Marel}, R.~P., \& {Alcock}, C. 2007, \apj, 668, 949
\bibitem[Besla et al.(2010)]{bes10}Besla, G.; Kallivayalil, N.; Hernquist, L.; van der Marel, R. P.; Cox, T. J.; Keres, D. ApJL, 721, 97
\bibitem[Besla et al.(2012)]{bes12}Besla, G., Kallivayalil, N., Hernquist, L., et al. 2012, 421, 2109 
\bibitem[Bessel \& Brett (1988)]{bb88}	Bessell, M. S., Brett, J. M., 1988, 100, 1134
\bibitem[Bono et al.(2005)]{bono05}Bono, G., Marconi, M., Cassisi, S., et al. 2005, ApJ, 621, 966
\bibitem[Caldwell \& Coulson (1986)]{cc86}Caldwell, J.A.R. \& Coulson, I.M. 1986, MNRAS, 218, 223
\bibitem[Cordier et al.(2003)]{cor03}Cordier, D., Goupil, M. J., Lebreton, Y., 2003, A\&A, 409, 491
\bibitem[{{Crowl} {et~al.}(2001){Crowl}, {Sarajedini}, {Piatti}, {Geisler},
  {Bica}, {Clari{\'a}}, \& {Santos}}]{c01}
{Crowl}, H.~H., {Sarajedini}, A., {Piatti}, A.~E., {Geisler}, D., {Bica}, E.,
  {Clari{\'a}}, J.~J., \& {Santos}, Jr., J.~F.~C. 2001, \aj, 122, 220
\bibitem[{{de Vaucouleurs} \& {Freeman}(1972)}]{df72}
{de Vaucouleurs}, G. \& {Freeman}, K.~C. 1972, Vistas in Astronomy, 14, 163
\bibitem[Diaz \& Bekki (2012)]{db12}Diaz, J. D., Bekki, K., 2012, ApJ, 750, 36	
\bibitem[Dobbie et al.(2014)]{do14}Dobbie, P. D., Cole, A. A., Subramaniam, A., Keller, S., 2014, MNRAS, 442, 1663
\bibitem[Efremov(1978)]{efr78}Efremov, Y. N. 1978, Soviet Astron., 22, 161
\bibitem[Efremov(2003)]{efr03} 2003, Astron. Rep., 47, 1000
\bibitem[Efremov \& Elmegreen (1998)]{efel98}Efremov, Y. N., \& Elmegreen, B. G. 1998, MNRAS, 299, 588
\bibitem[{{Evans} \& {Howarth}(2008)}]{eh08}
{Evans}, C.~J. \& {Howarth}, I.~D. 2008, \mnras, 386, 826
\bibitem[{{Gardiner} \& {Noguchi}(1996)}]{gn96}
{Gardiner}, L.~T. \& {Noguchi}, M. 1996, \mnras, 278, 191
\bibitem[Gieren et al. (1998)]{Gieren98}Gieren, W, P., Fouqué, P., Gómez, M., 1998, ApJ, 496, 17
\bibitem[Glatt et al.(2010)]{gla10}Glatt, K., Grebel, E. K., Koch, A., 2010, A\&A, 517, 50
\bibitem[Grebel \& Brandner (1998)]{grbr98}Grebel, E. K., \& Brandner, W. 1998, in The Magellanic Clouds and Other
Dwarf Galaxies, ed. T. Richtler \& J. M. Braun (Aachen: Shaker-Verlag), 151
\bibitem[{{Groenewegen}(2000)}]{g00}
{Groenewegen}, M.~A.~T. 2000, \aap, 363, 901
\bibitem[Haschke et al. (2012b)]{has12b}
	Haschke, R., Grebel, Eva K., Duffau, S., 2012, \aj, 144, 106
\bibitem[Haschke et al. (2012a)]{has12a}
	Haschke, R., Grebel, Eva K., Duffau, S., 2012, \aj, 144, 107
\bibitem[{{Harris} \& {Zaritsky}(2006)}]{hz06} {Harris}, J. \& {Zaritsky}, D., 2006, \aj, 131, 2514
\bibitem[Indu \& Subramaniam (2011)]{is11}Indu, G., Subramaniam, A., 2011, A\&A, 535, 115
\bibitem[Joshi et al.(2014)]{jj14}Joshi, Y. C., Joshi, S., 2014, New Astronomy, 28, 27
\bibitem[{{Kallivayalil} {et~al.}(2006{\natexlab{a}}){Kallivayalil}, {van der
  Marel}, \& {Alcock}}]{k06a}
{Kallivayalil}, N., {van der Marel}, R.~P., \& {Alcock}, C. 2006{\natexlab{a}},
  \apj, 652, 1213

\bibitem[{{Kallivayalil} {et~al.}(2006{\natexlab{b}}){Kallivayalil}, {van der
  Marel}, {Alcock}, {Axelrod}, {Cook}, {Drake}, \& {Geha}}]{k06b}
{Kallivayalil}, N., {van der Marel}, R.~P., {Alcock}, C., {Axelrod}, T.,
  {Cook}, K.~H., {Drake}, A.~J., \& {Geha}, M. 2006{\natexlab{b}}, \apj, 638,
  772
\bibitem[Kallivayalil et al.(2013)]{k13}Kallivayalil, N., van der Marel, R P., Besla, G., et al. 2013, ApJ, 764, 
161
\bibitem[Kanbur et al. (2006)]{kan06}Kanbur, Shashi M.; Ngeow, Chow-Choong., 2006, MNRAS, 369, 705
\bibitem[Kapakos et al. (2011)]{Kapakos11}Kapakos, E.; Hatzidimitriou, D.; Soszy{\~n}ski, I., 2011, MNRAS, 415, 1366
\bibitem[Laney \& Stobie (1986)]{ls86}Laney, C. D., Stobie, R. S., 1986, MNRAS, 222, 449L
\bibitem[Magnier et al. (1997)]{mg97}Magnier, E. A., Prins, S., Augusteijn, T., van Paradijs, J., \& Lewin, W. H. G.
1997, A\&A, 326, 442
\bibitem[Marconi et al. (2006)]{mar2006}Marconi, M., Bono, G., Caputo, F., Cassisi, S., Pietrukowicz, P., 
Pietrzynski, G., Gieren, W., 2006, MmSAI, 77, 67
\bibitem[Massey et al. (1995)]{Massey95}Massey, Philip; Lang, Cornelia C.; Degioia-Eastwood, Kathleen; Garmany, Catharine D., 
1995, ApJ, 438, 188
\bibitem[{{Mathewson} {et~al.}(1986){Mathewson}, {Ford}, \&
  {Visvanathan}}]{mf86}
{Mathewson}, D.~S., {Ford}, V.~L., \& {Visvanathan}, N. 1986, \apj, 301, 664

\bibitem[{{Mathewson} {et~al.}(1988){Mathewson}, {Ford}, \&
  {Visvanathan}}]{mf88}
---. 1988, \apj, 333, 617
\bibitem[Ngeow \& Kanbur (2006)]{nk06} Ngeow, Chow-Choong., Kanbur, Shashi M.; 2006, MNRAS, 369, 723
\bibitem[Nidever et al.(2013)]{nid13}Nidever, D. L., Monachesi, A., Bell, E. F., et al. 2013, ApJ, 779, 145
\bibitem[{{Nikolaev} {et~al.}(2004){Nikolaev}, {Drake}, {Keller}, {Cook},
  {Dalal}, {Griest}, {Welch}, \& {Kanbur}}]{n04}
{Nikolaev}, S., {Drake}, A.~J., {Keller}, S.~C., {Cook}, K.~H., {Dalal}, N.,
  {Griest}, K., {Welch}, D.~L., \& {Kanbur}, S.~M. 2004, \apj, 601, 260
\bibitem[Pietrzy{\~n}ski \& Udalski (2000)]{pu00}
	Pietrzy{\~n}ski, G., Udalski, A., 2000, AcA, 50, 337
\bibitem[{{Sandage} {et~al.}(2009){Sandage}, {Tammann}, \&
  {Reindl}}]{sandage09}
{Sandage}, A., {Tammann}, G.~A., \& {Reindl}, B. 2009, \aap, 493, 471
\bibitem[Schlegel et al.(1998)]{sch98}Schlegel, D. J., Finkbeiner, D. P., Davis, M., 1998, ApJ, 500, 525
\bibitem[Seth et al. (2005)]{Seth05}Seth, Anil C.; Dalcanton, Julianne J.; de Jong, Roelof S., AJ, 130, 1574
\bibitem[{{Stanimirovi{\'c}} {et~al.}(2004){Stanimirovi{\'c}},
  {Staveley-Smith}, \& {Jones}}]{stan04}
{Stanimirovi{\'c}}, S., {Staveley-Smith}, L., \& {Jones}, P.~A. 2004, \apj,
  604, 176
\bibitem[{{Sharpee} {et~al.}(2002){Sharpee}, {Stark}, {Pritzl}, {Smith},
  {Silbermann}, {Wilhelm}, \& {Walker}}]{sharpee02}
{Sharpee}, B., {Stark}, M., {Pritzl}, B., {Smith}, H., {Silbermann}, N.,
  {Wilhelm}, R., \& {Walker}, A. 2002, \aj, 123, 3216

\bibitem[{{Soszy{\~n}ski} {et~al.}(2010
){Soszy{\~n}ski},
  {Poleski}, {Udalski}, {Szyma{\~n}ski}, {Kubiak}, {Pietrzy{\~n}ski},
  {Wyrzykowski}, {Szewczyk}, \& {Ulaczyk}}]{u10smcc}
{Soszy{\~n}ski}, I., {Poleski}, R., {Udalski}, A., {Szyma{\~n}ski}, M.~K.,
  {Kubiak}, M., {Pietrzy{\~n}ski}, G., {Wyrzykowski}, {\L}., {Szewczyk}, O., \&
  {Ulaczyk}, K. 2010{\natexlab{a}}, \actaa, 60, 17
\bibitem[Subramanian \& Subramaniam (2012)]{ss12}Subramanian, S., Subramaniam, A., 2012, ApJ, 744, 128
\bibitem[Tammann et al.(2011)]{tam11}Tammann, G. A., Reindl, B., \& Sandage, A., 2011, A\&A, 531, 134
\bibitem[{{Udalski} {et~al.}(1999){Udalski}, {Szymanski}, {Kubiak},
  {Pietrzy{\~n}ski}, {Soszynski}, {Wozniak}, \& {Zebrun}}]{u99cep}
{Udalski}, A., {Szymanski}, M., {Kubiak}, M., {Pietrzynski}, G., {Soszynski},
  I., {Wozniak}, P., \& {Zebrun}, K. 1999, \actaa, 49, 201
\bibitem[{{van der Marel} \& {Cioni}(2001)}]{vc01}
{van der Marel}, R.~P. \& {Cioni}, M.-R.~L. 2001, \aj, 122, 1807
\bibitem[{{van der Marel} {et~al.}(2009){van der Marel}, {Kallivayalil}, \&
  {Besla}}]{v09}
{van der Marel}, R.~P., {Kallivayalil}, N., \& {Besla}, G. 2009, in IAU
  Symposium, Vol. 256, IAU Symposium, ed. {J.~T.~van Loon \& J.~M.~Oliveira},
  81--92
\bibitem[Valle et al.(2009)]{va09}Valle, G., Marconi, M., Degl$'$Innocenti, S., Prada Moroni, P. G., 2009, A\&A, 507, 1541
\bibitem[Weinberg \& Nikolaev (2001)]{wn01}Weinberg, M.D., \& Nikolaev, S, 2001, ApJ, 548, 712
\bibitem[{{Zaritsky} {et~al.}(2000){Zaritsky}, {Harris}, {Grebel}, \&
  {Thompson}}]{z00}
{Zaritsky}, D., {Harris}, J., {Grebel}, E.~K., \& {Thompson}, I.~B. 2000,
  \apjl, 534, L53
\end{thebibliography}
\end{document}